\pdfoutput=1
\documentclass[11pt,a4paper]{article}
\usepackage{authblk}
\usepackage{geometry}
\usepackage{xcolor}
\usepackage{amsmath,amssymb}
\numberwithin{equation}{section}
\usepackage{graphicx}
\usepackage{subcaption}
\usepackage[colorlinks=true,allcolors=blue]{hyperref}
\usepackage{cite}
\usepackage{comment}

\title{{\bf Black holes seeding cosmological phase transitions}}
\author[1,2]{Basem Kamal El-Menoufi\thanks{basem.el-menoufi@manchester.ac.uk}}
\author[1]{Stephan J. Huber\thanks{s.huber@sussex.ac.uk}}
\author[1]{Jonathan P. Manuel\thanks{j.manuel@sussex.ac.uk}}
\affil[1]{Department of Physics $\&$ Astronomy, 
University of Sussex, Falmer, Brighton, BN1 9QH, United Kingdom}
\affil[2]{Consortium for Fundamental Physics, School of Physics and Astronomy, University of Manchester, Manchester, M13 9PL, United Kingdom}
\date{}

\begin{document}

\maketitle
\begin{abstract}
We consider a generic first-order phase transition at finite temperature, and investigate to what extent a population of primordial black holes, of variable masses, can affect the rate of bubble nucleation. Using a thin-wall approximation, we construct the Euclidean configurations that describe transition at finite temperature. After the transition, the remnant black hole mass is dictated dynamically by the equations of motion. The transition exponent is computed, and displays an explicit dependence on temperature. We find the configuration with the lowest Euclidean action to be static and $O(3)$ symmetric; therefore, the transition takes place via thermal excitation. The transition exponent exhibits a strong dependence on the seed mass black hole, $M_+$, being {\em almost} directly proportional. A new nucleation condition in the presence of black holes is derived and the nucleation temperature is compared to the familiar flat-space result, i.e.~$S_3/T$. For an electroweak-like phase transition it is possible to enhance the nucleation rate if $M_+ \lesssim 10^{15} M_{\rm P}$. Finally, we outline the possible transition scenarios and the consequences for the power spectrum of stochastic gravitational waves produced due to the first-order phase transition.
\end{abstract}

\section{Introduction}

Cosmological phase transitions at different epochs may have played a major role in the history of the Universe, potentially occurring anywhere between the QCD ($\sim 100\, \textrm{MeV}$) and GUT scales ($\sim 10^{16}\, \textrm{GeV}$)~\cite{Mazumdar:2018dfl}. A much studied example is the electroweak phase transition, due to which known elementary particles acquired their masses. Of particular interest are first-order phase transitions which proceed through the nucleation and expansion of bubbles; the two phases are separated by a bubble wall, inside exists the new ``true" vacuum while outside the old ``false" vacuum. In particle physics, there exists a continuous interest in first-order transitions because they provide departure from thermal equilibrium, as required in the process of baryogengesis~\cite{Kuzmin:1985mm,Shaposhnikov:1986jp,Shaposhnikov:1987tw,Morrissey:2012db,Konstandin:2013caa}. In addition, the collision of expanding bubbles yield a stochastic background of gravitational waves that is well within the reach of LISA~\cite{Hogan:1986qda,Kosowsky:1992rz,Kamionkowski:1993fg,Caprini:2015zlo,Caprini:2019egz}. 

The first description of vacuum decay in continuum field theory was famously given by Coleman and Callan~\cite{Coleman:1977py,Callan:1977pt} and later extended by Linde~\cite{Linde:1981zj} to the case of finite-temperature phase transitions. Therein it was determined that the probability for nucleation of the new phase, per unit time and per unit volume, is given by
\begin{equation}\label{eqn:nuc_rate}
    \frac{\Gamma}{V} = Ae^{-B}\ ,
\end{equation}
where $A$ is a coefficient of mass dimension four and the tunneling exponent, $B$, is the difference between the Euclidean action of the tunneling configuration and that of the false vacuum. Due to the exponential dependence, $B$ is the quantity mostly studied, while an estimate of $A$ usually suffices. On one hand, vacuum decay is attributed to quantum tunneling, and therefore is appropriate for zero-temperature phase transitions. On the other hand, at finite temperature two distinct physical effects exist; the phase transition could proceed either via ``thermally-assisted" quantum tunneling or by classical thermal excitation over the potential barrier.

Later on, Coleman and de Luccia (CdL) raised the question about the effect of gravitation on the dynamics of vacuum decay~\cite{Coleman:1980aw}. Assuming $O(4)$ symmetry and using the thin-wall approximation, an appropriate limit when the reduction in vacuum energy is small compared to the height of the barrier, they found simple yet stark results. If one tunnels from a space with positive vacuum energy to a smaller, yet still positive, or zero vacuum energy then gravitation makes vacuum decay more probable. On the other hand, if one tunnels from a space with zero or negative vacuum energy then gravitation makes vacuum decay less probable. While this indeed comprised an important insight, effects of gravity remained purely academic for phenomenology. Apart from negligible corrections, the nucleation rate remained essentially unchanged from flat-space given any practical values of the surface tension and vacuum energy~\cite{Coleman:1980aw}.

Notwithstanding, this might not be end for gravity. In particular, can primordial black holes (PBHs) influence cosmological phase transitions? In recent years, the interest in PBHs has rapidly intensified, see the review articles~\cite{Sasaki:2018dmp,Carr:2016drx,Carr:2020gox}, and it seems inevitable that we revisit the dynamics of phase transitions in the presence of PBHs. The first study about the topic of black holes and phase transitions was done by Hiscock~\cite{Hiscock:1987hn}, and was concerned with vacuum transitions at zero temperature. He found that black holes {\em do} indeed increase the nucleation probability. Precisely, the presence of black holes could cause the Euclidean action to diminish by up to a factor of approximately two. However, the analysis was limited in a number of ways. Firstly, the black hole mass was kept constant throughout the transition. Secondly, conical singularities, arising from an unmatched Euclidean time period and inverse Hawking temperature, were not accounted for.

Recently, Gregory {\em et. al.}~\cite{Gregory:2013hja,Burda:2015isa,Burda:2015yfa,Burda:2016mou} undertook more work in this direction that overcame the shortcomings of~\cite{Hiscock:1987hn}, in particular, the treatment of conical singularities. Therein, the focus was mainly on the Higgs vacuum (in)-stability in the presence of black holes\footnote{The stability of the electro-weak vacuum, in pure de Sitter space, has been investigated in Ref.~\cite{Rajantie:2016hkj}.}, and they discovered that the Euclidean action could be arbitrarily reduced depending on the seed black hole mass. The central construction is based on the thin-wall approximation~\cite{Coleman:1977py}, and utilizes Israel's junction conditions~\cite{Israel:1966rt} to smoothly glue the spacetimes that represent the two phases. Each vacuum state contains a black hole, and is therefore given by a static Schwarzschild-de-Sitter (SdS) black hole. Applying the junction conditions then yields a dynamical equation for the ``bubble" wall, whose solution determines the full instanton. We also note that further studies have been undertaken in Refs.~\cite{Canko:2017ebb,Gorbunov:2017fhq,Mukaida:2017bgd,Oshita:2018ptr} along complementary directions.

In the standard setting of early Universe cosmology, there is an epoch of inflation followed by reheating. The Universe then reaches a state of thermal equilibrium, after which phase transitions are likely to proceed via thermal, rather than quantum, fluctuations. In most scenarios, the phase transition proceeds thermally via finite-temperature effects in the potential. For first-order phase transitions, as the Universe cools down the potential develops a barrier that separates two phases, thereby initiating the phase transition. In addition, any cosmological first-order phase transition will, through the expanding and colliding bubbles of the new phase, produce a stochastic gravitational wave (GW) background. Of particular interest are transitions happening around the electroweak scale as the expected signal of GW's is within the sensitivity of LISA\cite{Caprini:2015zlo,Caprini:2019egz}.

In this paper we investigate the potential effects of primordial black holes on the nucleation rate of a generic first-order phase transition that proceeds thermally. This situation is more complicated than the vacuum case, as it is {\em not} obvious a priori how to define the appropriate finite-temperature instanton, that should ultimately be used to compute the Euclidean action. In particular, which solutions of the equation of motion, describing the bubble wall motion, are relevant for the phase transition? We present a prescription that singles out the relevant tunneling configurations, the details of which are presented in Sect.~\ref{conds}. Our prescription is entirely guided by the analogy with standard finite-temperature tunneling in flat-space, which is reviewed in Sect.~\ref{qfttun}. Our method offers a consistent formalism to quantify the effects of PBHs on any generic cosmological first-order phase transition, at least in the thin-wall regime. This sheds new light on the features of the GW spectrum generated by the phase transition, and paves the way to study possible links with the properties of PBHs.

The paper is structured as follows. In Sect.~\ref{sec:QFT} we explicitly describe the methods of tunneling, first in quantum mechanics before extending to quantum field theory. There are three generic mechanisms: zero temperature quantum tunneling through infinite-period bounces, thermally assisted quantum tunneling through finite-period bounces, and thermal excitation through static bubbles. Black holes are introduced in Sect.~\ref{sec:BH_and_PT} where we begin by recapping the Israel thin-wall formalism developed by Refs.~\cite{Gregory:2013hja,Burda:2015yfa} before moving on to quantitatively analysing the solutions and how they are applied to finite temperature cosmological phase transitions. In Sect.~\ref{sec:bounce_action} we derive the bounce action, determine a new nucleation criteria and then apply our approach to the example of a first-order EWPT. We outline the pathway to making phenomenological predictions in Sect.~\ref{sec:pheno}. Sect.~\ref{sec:discussion} contemplates the various scenarios and discusses the consequences for gravitational waves. A summary is then given in Sect.~\ref{sec:summary}.

\section{Bubbles vs Bounces in QFT}\label{sec:QFT}

\subsection{Tunneling in Quantum Mechanics}
\begin{figure}[t]
     \centering
     \begin{subfigure}{0.49\textwidth}
     \includegraphics[scale=0.50]{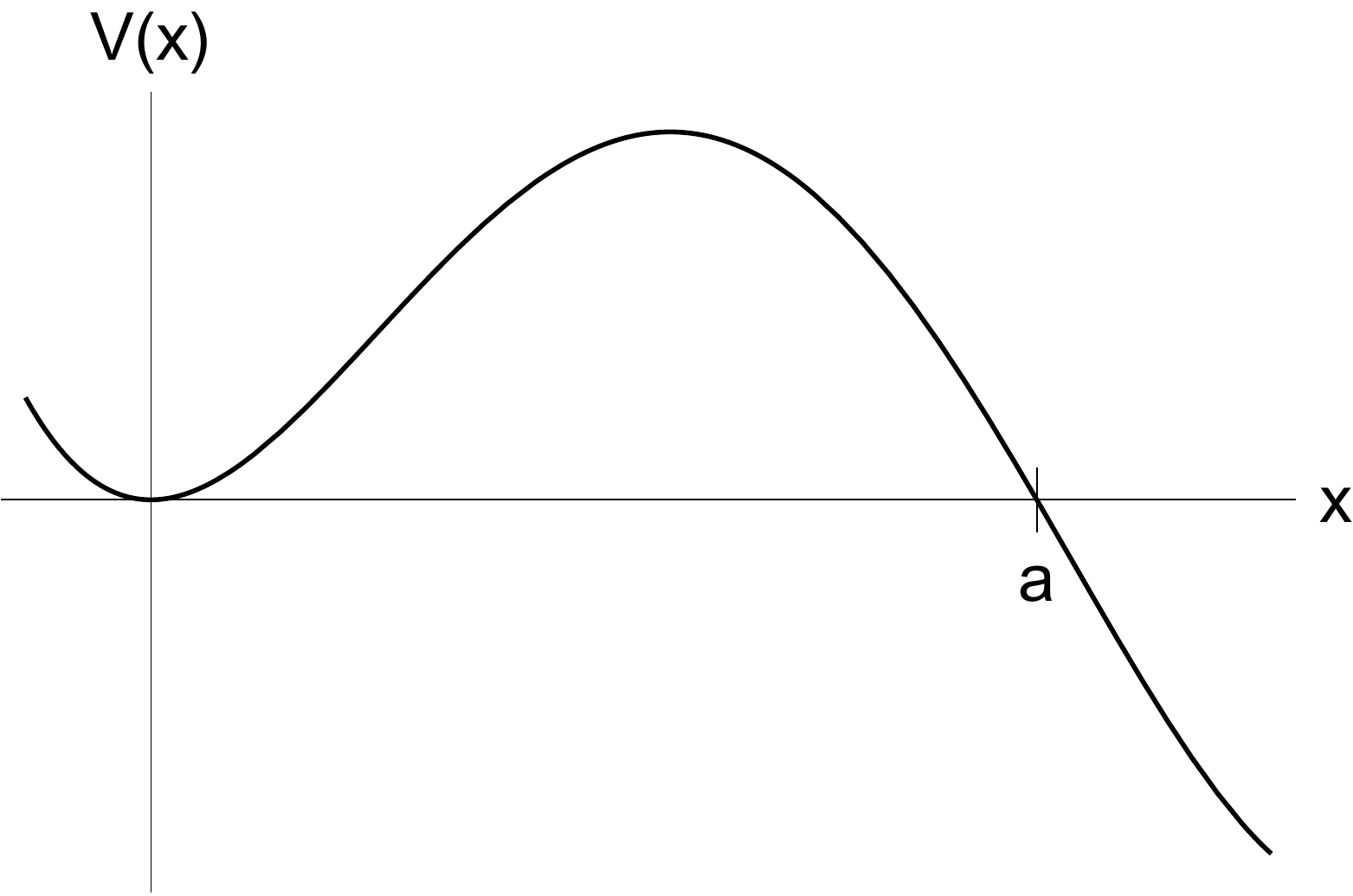}
     \end{subfigure}
     \hfill
     \begin{subfigure}{0.49\textwidth}
     \includegraphics[scale=0.50]{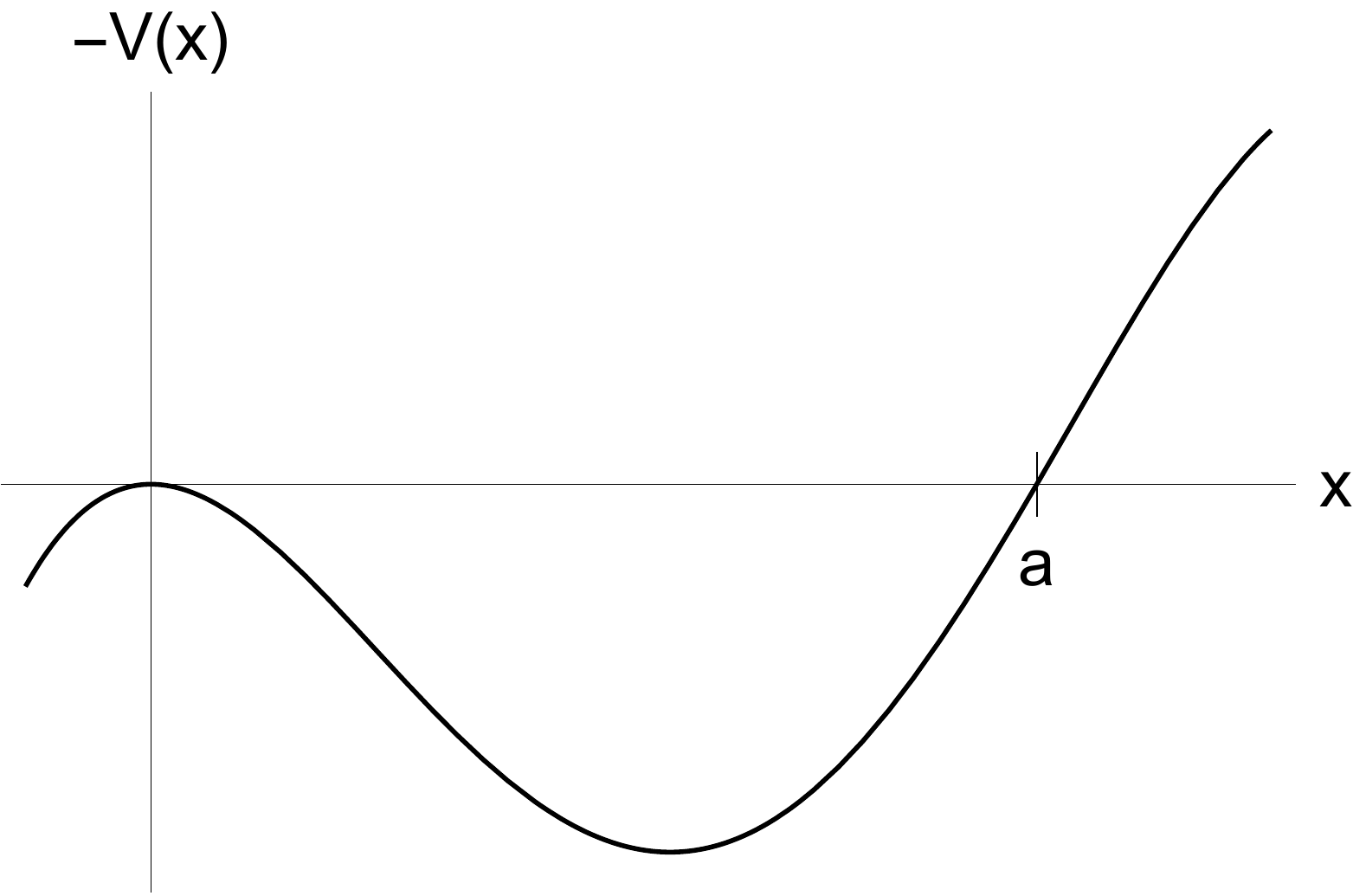}
     \end{subfigure}
     \caption{Left: A typical tunneling potential $V(x)$. Right: The inverted tunneling potential $-V(x)$.}
     \label{fig:tun_pot}
\end{figure}

Our knowledge about tunneling in quantum field theory comes directly from non-relativistic quantum mechanics, in particular, single particle systems. The typical potential, $V(x)$, one is interested in is shown in Fig.~\ref{fig:tun_pot}, where a particle is initially localized to the left of the barrier at the bottom of the potential well. Quantum mechanics renders such a state unstable and it becomes paramount to compute the lifetime of unstable states. The WKB approximation offers an analytic technique to study quantum tunneling for a generic potential. The finite probability, per unit time, for the particle to quantum tunnel through the barrier is given by
\begin{align}\label{1prate}
\Gamma \simeq A e^{-B}, \quad B := 2 \int_0^a \sqrt{2V(x)} dx \ \ .
\end{align}
where $A$ is a prefactor with the dimensions of energy. There is a remarkable feature of the tunneling exponent in that it directly derives from a special solution to the Euclidean equation of motion of the system, i.e.
\begin{align}\label{1peuceom}
m \frac{d^2x}{d\tau^2} + \frac{\left(-V(x)\right)}{dx} = 0 \ \ ,
\end{align}
where $\tau$ is Euclidean time. Eq.~\eqref{1peuceom} simply describes a classical particle of mass $m$ moving in the inverted potential of Fig.~\ref{fig:tun_pot}. 
Coleman~\cite{Coleman:1977py} observed that Eq.~(\ref{1peuceom}) admits periodic solutions known as \emph{bounces}. To see this, we write the conservation of Euclidean energy
\begin{align}\label{1peuccons}
\frac12 m^2 \left(\frac{dx}{d\tau}\right)^2 + \left(-V(x)\right)  =0 \ \ ,
\end{align}
so indeed the particle can start at the origin, slide down to reach $x=a$ and then bounce back to the origin. In particular, inspection of Eq.~(\ref{1peuccons}) reveals that the particle can only reach the origin as $\tau \to \pm \infty$ which demonstrates that the bounce has an infinite period. Hence, the bounce solution, $x_\infty(\tau)$, has the following boundary conditions
\begin{align}\label{1pbcs}
\lim_{\tau \to \pm \infty} x_\infty(\tau) = 0, \quad x_\infty(0) = a, \quad \frac{dx_\infty(0)}{d\tau} = 0 \ \ .
\end{align}
If we now insert the bounce solutions in the Euclidean action, one recovers the exponent $B$ in Eq.~(\ref{1prate}),
\begin{align}
S_{\text{E}} = \int_{-\infty}^\infty d\tau \left(\frac12 m^2 \left(\frac{dx_b}{d\tau}\right)^2 + V(x_b) \right) = B \ \ .
\end{align}

In fact this is not the end of story; another class of Euclidean solutions play a dominant role in describing the decay of {\em thermally} excited states. These solutions also represent bounce-like behavior, but with a finite period in Euclidean time. Let us start by recalling the tunneling exponent of a thermally excited state, with energy $E$ and inverse temperature $\beta$, 
\begin{align}\label{1pBthermal}
B(E;\beta) = \beta E  + 2  \int_{x_1}^{x_2} \sqrt{2m (V(x) - E)} dx \ \ ,
\end{align}
where the first factor is the Boltzman suppression and $(x_1,x_2)$ are the classical turning points, which we remind are functions of energy. Notice that the energy is kept arbitrary at this stage, but to find the appropriate decay rate in Eq.~(\ref{1prate}) one has to minimize the exponent in Eq.~\eqref{1pBthermal}, at fixed temperature, with respect to energy. One gets
\begin{align}
\frac{\partial B}{\partial E}(E_0) = 0  \ \  \Rightarrow  \ \ \beta = 2 m \int_{x_1}^{x_2} dx \frac{1}{ \sqrt{2m (V(x) - E_0)}} \ \ ,
\end{align}
which then determines the energy as a function of temperature. Plugging $E_0$ back into Eq.~\eqref{1pBthermal} yields the decay rate exponent. The latter can be derived from another class of solutions to Eq.~(\ref{1peuceom}), i.e. those solutions with finite energy
\begin{align}
\frac12 m^2 \left(\frac{dx}{d\tau}\right)^2 + \left(-V(x)\right)  = - E \ \ .
\end{align}
Clearly a solution to Eq.~(\ref{1peuceom}) with finite energy $E$ is also a bounce, albeit with a finite period given by
\begin{align}\label{1pperiod}
 P(E) = 2 m \int_{x_1}^{x_2} dx \frac{1}{ \sqrt{2m (V(x) - E)}} \ \ .
\end{align} 
Now Eq.~(\ref{1pperiod}) yields the energy that satisfies $P(E_0 )=\beta$. It is straightforward to then obtain the finite-period bounce that we denote by $x_\beta(\tau)$. Finally, we evaluate the Euclidean action on this periodic bounce to recover Eq.~(\ref{1pBthermal})
\begin{align}
S_{\text{E}} = \int_{-\beta/2}^{\beta/2} d\tau \left(\frac12 m^2 \left(\frac{dx_\beta}{d\tau}\right)^2 + V(x_\beta) \right) = B(E_0;\beta) \ \ .
\end{align}
As one might expect, if the temperature is high enough the particle gets excited to the top of the barrier and, therefore, {\em classically} transitions to the allowed region. The temperature at which this takes place can be estimated by approximating the inverted potential, at its minimum, as $V(x_0) \simeq V_0 - \frac12 m \omega^2_0 x^2$. A particle moving in the inverted potential, near $x=x_0$, then experiences a {\em fixed} period $2\pi/\omega_0$. Therefore, Eq.~(\ref{1peuceom}) possesses no finite-period Euclidean solutions when the temperature is such that
\begin{align}
\beta < \beta_0 := \frac{2\pi}{\omega_0}  \ \ ,
\end{align}
and the unique available solution becomes the static configuration $x_{\text{s}}(\tau) = x_0$ with a tunneling exponent given by
\begin{align}
S_{\text{E}} = \beta V_0  \ \ .
\end{align}

To summarize, we have three separate solutions to the Euclidean equation of motion and each describe a distinct physical situation. First, we have the infinite-period bounce $x_\infty(\tau)$ which describes the decay of the vacuum state. Second, the finite-period bounce $x_\beta(\tau)$ describes the decay of a thermally excited state. Third, the static solution $x_{\text{s}}$ describes the classical excitation of the particle over the potential barrier.

\subsection{Tunneling in Quantum Field Theory}\label{qfttun}
The close connection between Euclidean solutions and tunneling exponents is pivotal for quantum field theory. If the potential functional in the quantum field theory exhibits a barrier, one can mimic the strategy drawn from quantum mechanics to compute the tunneling probability per unit time and per unit volume. Here we clearly need to understand what kind of boundary conditions one has to impose on Euclidean solutions that describe the tunneling process. We write down the Euclidean equations of motion and then proceed to find bounce, as well as static, solutions. The Euclidean action evaluated on the these solutions is then interpreted as providing the decay exponent in Eq.~(\ref{1prate}). The justification of this procedure in QFT is best offered by the work of Coleman and Callan~\cite{Coleman:1977py,Callan:1977pt}, who reformulated the tunneling problem in quantum mechanics using Euclidean path integral methods.

In summary, we have three physical scenarios echoing the story in quantum mechanics. The only new input concerns the {\em spatial} boundary conditions imposed on the Euclidean solutions. As the solutions become extended in 3D space, we need to ensure that the action remains finite. 

\paragraph{Infinite-period bounces \& vacuum decay:}

First, we have a QFT of a scalar field, $\varphi$, held at zero-temperature whose Euclidean equation of motion reads
\begin{align}\label{phieomzeroT}
\frac{d^2\varphi}{d\tau^2} + \nabla^2 \varphi - \frac{dV\left(\varphi\right)}{d\varphi} = 0  \ \ ,
\end{align}
where the potential $V(\varphi)$ has a barrier separating the two vacua. The bounce solution is fully symmetric in Euclidean time, and thus one can focus on the semi-infinite interval, $\tau \in [0,\infty)$. The temporal boundary conditions are exactly identical to Eq.~\eqref{1pbcs}
\begin{align}\label{phibczeroT}
\lim_{\tau \to + \infty} \varphi(\tau,\vec{x}) = \varphi_+, \quad \frac{d\varphi(0,\vec{x})}{d\tau}=0 \ \ ,
\end{align}
and to ensure finiteness of the Euclidean action we further impose a spatial boundary condition
\begin{align}\label{phibc2zeroT}
 \lim_{|\vec{x}| \to \infty} \varphi(\tau,\vec{x}) = \varphi_+ \ \ .
 \end{align}

Indeed, one has to resort to numerical techniques to solve this system. Nevertheless, Coleman proved that the solution with the lowest action is $O(4)$ invariant~\cite{Coleman:1977py}, a fact that simplifies the situation considerably. Now Eqs.~\eqref{phieomzeroT}-\eqref{phibc2zeroT} become
\begin{align}\label{phieomO4}
\frac{d^2\varphi}{d\rho^2} + \frac{3}{\rho} \frac{d\varphi}{d\rho}  - \frac{dV\left(\varphi\right)}{d\varphi} = 0   \ \ , \quad \lim_{\rho \to \infty} \varphi(\rho) = \varphi_+ \ \ , \quad \frac{d\varphi(0)}{d\rho}= 0 \ \ ,
\end{align}
where the 4D radius is $\rho^2 = \tau^2 + \vec{x}^2$. The last condition in the above equation is to ensure the solution is regular at the origin $\rho = 0$. Notice that since the $O(4)$ solution is even in $\tau$, the second condition in Eq.~(\ref{phibczeroT}) is automatically satisfied. Eq.~\eqref{phieomO4} presents an ODE, which can be solved by numerical methods. 

\paragraph{Finite-period bounces \& tunneling at finite temperature:} 

\begin{figure}[t]
    \centering
    \includegraphics[width=0.4\textwidth]{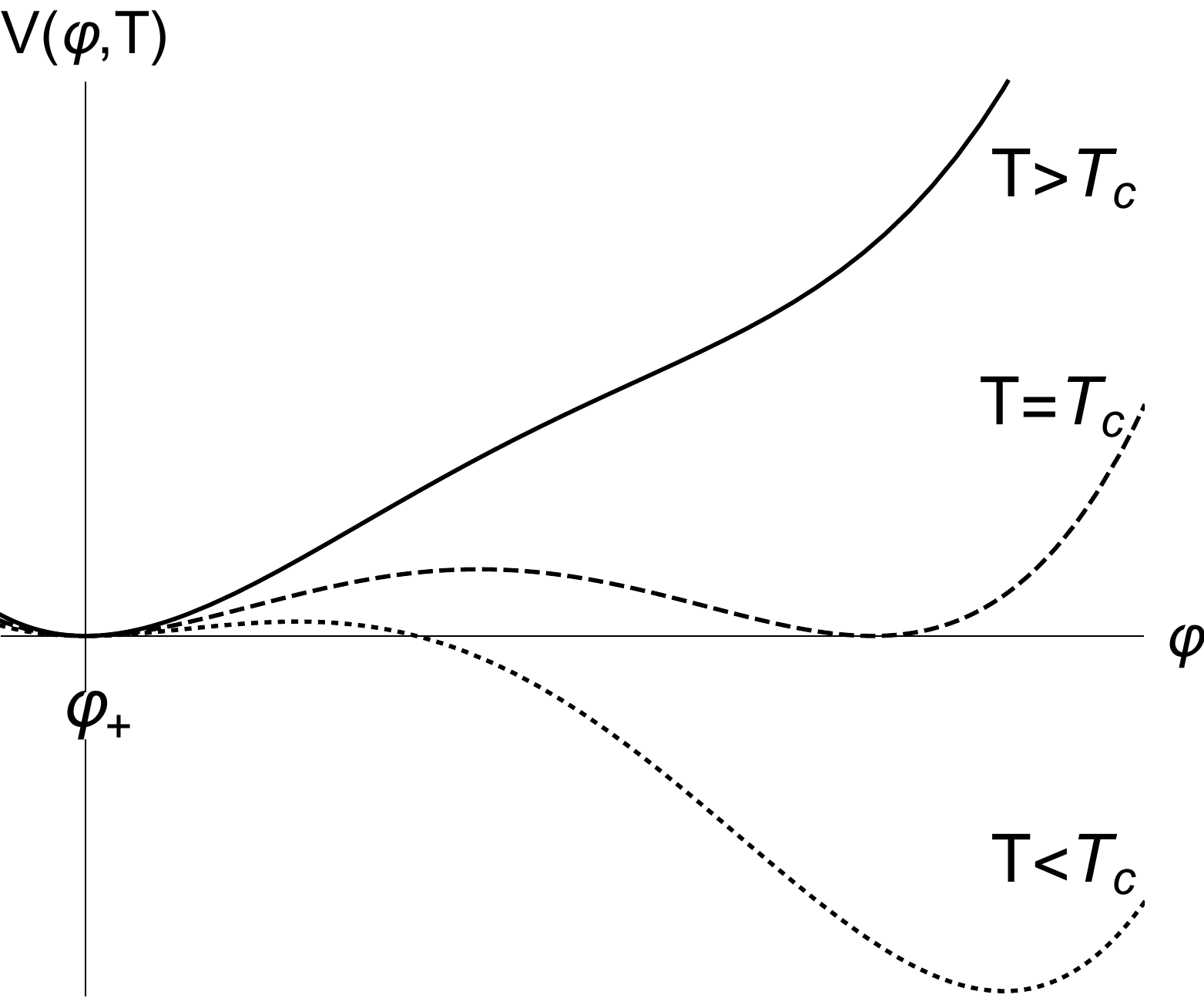}
    \caption{An example scalar field potential where the barrier is generated through finite temperature effects. At large temperatures there is a single minimum at $\varphi_+$, the false vacuum. As the temperature decreases a second minimum forms at $\varphi_-$, the true vacuum, the location of which is temperature dependent. At the critical temperature $T_c$ the two minima are degenerate. Below the critical temperature the true vacuum is energetically favourable and a transition can occur.}
    \label{fig:scalar_pot_FT}
\end{figure}

Second, we have the situation relevant for cosmological phase transitions, whereby the theory is held at finite temperature, $\beta=1/T$, and the thermal potential, $V(\varphi,T)$, develops a barrier as the Universe cools down, see Fig.~\ref{fig:scalar_pot_FT}.
Therefore, the finite-temperature bounce satisfies 
\begin{align}\label{phieomT}
\frac{d^2\varphi}{d\tau^2} + \nabla^2 \varphi - \frac{dV\left(\varphi,T\right)}{d\varphi} = 0  \ \ ,
\end{align}
with the following boundary conditions
\begin{align}\label{phibcT1}
 \frac{d\varphi(\tau,\vec{x})}{d\tau}\bigg|_{\tau=0}=0\ , \quad \lim_{|\vec{x}| \to \infty} \varphi(\tau,\vec{x}) = \varphi_+\ , \quad  \frac{d\varphi(\tau,\vec{x})}{dr}\bigg|_{r=0}=0 \ \ .
\end{align}
The first condition clearly signifies the bouncing behavior, the second guarantees the Euclidean action is finite and the third assures the solution is regular at the origin. This set of conditions is not enough to guarantee a solution to Eq.~\eqref{phieomT}, in other words, we need an extra condition that describes the behavior of the finite-temperature bounce as $\tau \to \pm \beta/2$. We observe that the (conserved) Euclidean Hamiltonian, for a finite-period bounce, has to be non-zero given the thermal excitation of the system. Therefore, in contrast to vacuum decay described above, a finite-temperature solution can {\em not} approach the false vacuum at $\pm \beta/2$. Hence, the remaining condition must instead be on the velocity of the field, i.e.
\begin{align}\label{phibcT2}
\lim_{\tau \to \pm \beta/2} \ \frac{d\varphi(\tau,\vec{x})}{d\tau} = 0 \ \ ,
\end{align}
which manifestly describes the bouncing behavior of the solution. As a final remark, these solutions are almost never discussed in the literature, yet, we believe the conditions in Eqs.~\eqref{phibcT1} and \eqref{phibcT2} render the problem well-posed although it is not possible to make a concrete statement regarding whether a non-trivial solution exists. This can only be verified by explicit numerical methods.

\paragraph{Static solutions at high temperature:} 

Lastly, we have the familiar static solution at high temperatures. The equation of motion and boundary conditions are identical to the $O(4)$-symmetric bounce, except the dynamics take place in three dimensions. The solution with the minimum action is $O(3)$ invariant, i.e.
\begin{align}
\frac{d^2\varphi}{dr^2} + \frac{2}{r} \frac{d\varphi}{dr}  - \frac{dV\left(\varphi,T\right)}{d\varphi} = 0   \ \ , \quad \lim_{r \to \infty} \varphi(r) = \varphi_+ \ \ , \quad \frac{d\varphi(0)}{dr}= 0 \ \ ,
\end{align}
and clearly has a vanishing period. Although it is not possible in QFT to easily estimate the temperature at which the static solution dominates over the finite-period bounce, one can numerically compute the Euclidean action of both solutions as a function of temperature and utilize the smaller action as the decay exponent. 

\section{Black Holes and Cosmological Phase Transitions}\label{sec:BH_and_PT}

In this section we consider a scalar field theory at finite temperature propagating in a background spacetime that contains a black hole. In particular, we focus on the typical situation that a potential barrier is generated by finite temperature effects as shown in Fig.~\ref{fig:scalar_pot_FT}. Although the presentation is quite general, we have in mind an electroweak-like phase transition. As the Universe cools down, the scalar field eventually tunnels and we wish to compute the tunneling exponent in the presence of a primordial population of {\em static} black holes. We will conduct our study with one caveat, which concerns the contribution of the thermal plasma to the equations of motion. It is well known EW-scale phase transitions occur during radiation domination. As we set to solve the equations of motion we will ignore the contribution of the thermal plasma to the energy-momentum tensor of the system. We do so for two main reasons. First, the thin-wall approximation, used throughout, requires knowing the analytic solutions for the spacetime metric in both vacuum states. However, we are not aware of any closed form solutions describing a black hole in a Friedmann–Lema\^{\i}tre–Robertson–Walker (FLRW) Universe. Second, our main goal is to set up the appropriate formalism to compute the tunneling exponent for thermal transitions seeded by black holes. In particular, this enables us to conclude the dependence of the tunneling rate on both the seed mass and the temperature of the system. We do not expect the conclusions from our study to change significantly once we include the expansion of the Universe in the story. Technically, the caveats just mentioned comprise an approximation that could be justified on the ground that the Hubble time is the same order of magnitude as the typical lifetime of EW-like phase transitions. Finally, we assume a bare positive cosmological constant in the gravitational sector to allow the Universe, post-transition, to retain a positive vacuum energy that we try and keep close to the scale of dark energy. Altogether, we aim to include the effect of the Universe expansion on the tunneling process in future work.

\subsection{Thin-wall instantons}\label{thinwall}

The goal is to solve the Euclidean equations of motion of the coupled scalar-gravity system. The analysis is considerably simplified if we adopt the thin-wall approximation. In the absence of gravity, the approximation is valid as long as the radius of the bubble is large compared the Compton wavelength of the field \cite{Coleman:1977py}. In the presence of black holes, however, does the same criterion validate the approximation? We will show below that this is fortunately the case.

The Euclidean equations of motion are those of a scalar field, with a finite-temperature potential, minimally coupled to general relativity
\begin{align}\label{eombasic}
\nonumber
&G_{\mu\nu} + \Lambda_0 g_{\mu\nu} = \frac{1}{M_\textrm{P}^2} T_{\mu\nu} \ \ , \\\nonumber
&T_{\mu\nu}  = \partial_\mu \varphi \partial_\nu \varphi - g_{\mu\nu} \left( \frac12 g^{\alpha\beta} \partial_\alpha \varphi \partial_\beta \varphi + V(\varphi, T) \right)  \ \ , \\
&\nabla^\mu T_{\mu\nu} = 0 \quad  \Rightarrow  \quad \Box \varphi  - \frac{dV\left(\varphi,T\right)}{d\varphi} = 0   \ \ ,
\end{align}
where $\Box = g^{\mu\nu} \nabla_\mu \nabla_\nu$ is the covariant Laplacian and $G_{\mu\nu}$ is the Einstein tensor\footnote{Our convention for the Riemann tensor is that of Wald \cite{Wald:1984rg}.}. The equilibrium solutions are those with a homogeneous field profile permeating the most general static spherically symmetric spacetime, hence the metric has the general form
\begin{equation}\label{generalmetric}
ds^2 = f(r) d\tau^2 + \frac{dr^2}{f(r)} + r^2 \left(d\theta^2 + \sin^2 \theta d\phi^2 \right), \quad \varphi(\tau,\vec{x}) = \varphi_0(T)  \ \ .
\end{equation}
In the case of interest, where the spacetime contains a black hole, Euclidean time is periodic, i.e. $0 \leq \tau \leq \beta_\tau $. Nevertheless, we do not assume any restrictions on the period of Euclidean time, which renders the spacetime singular as it contains a conical singularity. The contribution of the conical singularity to the tunneling exponent will be computed according to the procedure given in \cite{Fursaev:1995ef}. The effective cosmological constant sourcing the spacetime is given by the {\em total} vacuum energy
\begin{equation}
\Lambda = \Lambda_0 + \frac{V(\varphi_0(T), T)}{M_\textrm{P}^2} \ \ ,
\end{equation}
and it proves useful to measure the cosmological constant in units of $M_{\text{P}}^2$, thus we introduce 
\begin{equation}
    \epsilon_\Lambda \equiv M_\textrm{P}^2 \Lambda \ \ .
\end{equation}

\noindent Now we wish to construct a thin-wall instanton that interpolates between the true and false vacua, given the typical scalar field finite-temperature potential in Fig.~\ref{fig:scalar_pot_FT}. The analytic expression for such a potential will be given later on when we construct instantons for electroweak-like phase transitions. Specifically, inside the wall we have 
\begin{equation}
ds_{-}^2=f(r_-)d\tau_-^2 + f(r_-)^{-1}dr_-^2 + r_-^2 d\Omega_{2}^2, \quad \varphi = \varphi_0(T) \ \ ,
\end{equation}
while outside
\begin{equation}
ds_{+}^2=f(r_+)d\tau_+^2 + f(r_+)^{-1}dr_+^2 + r_+^2 d\Omega_{2}^2, \quad \varphi = 0 \ \ .
\end{equation}
At this stage the Euclidean periodicities ($\beta_{\tau_+}, \beta_{\tau_-}$) are arbitrary, and in general $\beta_{\tau_+} \neq \beta_{\tau_-}$. In the thin-wall approximation, the bubble wall is a hypersurface (thin layer) separating the two equilibrium vacua \cite{Hiscock:1987hn}. Spherical symmetry forces the induced metric on the wall to have the form
\begin{equation}\label{wallmetric}
ds^2_{\text{wall}} = d\lambda^2 + R^2(\lambda)d\Omega_{2}^2 \ \ ,
\end{equation}
where $\lambda$ is the proper time measured by a co-moving observer with the wall and $d\Omega_{2}^2$ is the standard metric on a unit $S^2$. Indeed, $\lambda$ is periodic, i.e. $0 \leq \lambda \leq \beta_\lambda$, which means the wall has topology of $S^2 \times S^1$. Notice that we have three distinct Euclidean times which, a priori, each have a unique period
\begin{align}
\beta_\lambda \neq \beta_{\tau_+} \neq \beta_{\tau_-} \ \ ,
\end{align} 
and only the dynamics will dictate any possible relation between these periods. To understand the geometry of the bubble wall, we introduce tangent vectors that span each side of the wall surface
\begin{equation}
e^\mu_{(1)\pm} = \dot{\tau}_\pm \frac{\partial}{\partial \tau_\pm} + \dot{R} \frac{\partial}{\partial r_\pm} , \quad e^\mu_{(2)} =\frac{\partial}{\partial \theta}, \quad e^\mu_{(3)} = \frac{\partial}{\partial \phi} \ \ ,
\end{equation}
where an over-dot denotes differentiation with respect to $\lambda$. We can immediately work out the first junction condition \cite{Israel:1966rt}, which requires the induced metrics on both sides of the wall to be identical. 
At the wall we must have $r_+ = r_- = R(\lambda)$. Additionally, we must have
\begin{equation}
    g_{\mu\nu} e^\mu_{(1)\pm} e^\nu_{(1)\pm}  = 1\ \ ,
\end{equation}
which yields
\begin{equation}
    f_\pm(R) \dot{\tau}_\pm^2 + \frac{1}{f_\pm(R)} \dot{R}^2 = 1 \label{eqn:juncone}   \ \ .
\end{equation}
Finally, we have the normal one-forms
\begin{equation}
 n_\mu^\pm = - \dot{R} d\tau_{\pm} + \dot{\tau}_\pm dr_\pm\ \ ,
\end{equation}
which are unit normalized by virtue of Eq.~(\ref{eqn:juncone}), and point outward from the wall surface. The second junction condition requires the jump in the extrinsic curvature to be proportional to the surface tension of the wall as follows \cite{Israel:1966rt}
\begin{equation}\label{secondjunc}
\Delta K_{ab} = - 8\pi G \left(S_{ab} - \frac{1}{2} S h_{ab}\right), \quad K_{ab,\pm} = e_{(a)} ^\mu  e_{(a)} ^\nu \nabla_{\mu} n_{\nu,\pm}\ \ ,
\end{equation}
where $h_{ab}$ is the first fundamental form or simply the induced metric in Eq.~(\ref{wallmetric}), $S_{ab}$ is the energy-momentum tensor of the wall, and $a$ and $b$ run over $(\lambda,\theta,\phi)$. As the scalar field tunnels through the barrier, and in the thin-wall approximation, the field gradients through the wall create a surface tension which in turn sources the wall geometry in Eq.~\eqref{secondjunc}. For a spherical wall with surface tension $\sigma$ we have \cite{Hiscock:1987hn}
\begin{align}
S_{ab} = - \sigma \, h_{ab}  \ \ ,
\end{align}
where $\sigma$ is typically evaluated at the critical temperature of the transition and, therefore, fully depends on the particular structure of the physics model.
Using the $(\theta$-$\theta)$ components of Eq.~(\ref{secondjunc}) we find
\begin{equation}\label{eqn:ex_curv_diff_comp}
\frac{1}{R} \left(f_+(R) \dot{\tau}_+ - f_-(R) \dot{\tau}_- \right) = -4 \pi G \sigma \ \ ,
\end{equation}
and, through Eq.~(\ref{eqn:juncone}), we can obtain an equation for $\dot{R}$
\begin{equation}
 \sqrt{f_+(R) - \dot{R}^2} - \sqrt{f_-(R) - \dot{R}^2} = -4 \pi G \sigma R \ \ .
\end{equation}
The above equation is not very useful due to the square root structure, and thus we can square twice to obtain 
\begin{align}\label{eqn:general_eom}
\dot{R}^2 = \frac{f_++f_-}{2}-\frac{(f_+-f_-)^2}{64\pi^2 G^2 \sigma^2 R^2} - 4\pi^2G^2\sigma^2 R^2 \ \ ,
\end{align}
which is our final equation of motion describing the bubble wall motion \cite{Hiscock:1987hn,Gregory:2013hja}. 

\subsection{Interlude: the Coleman-de Lucia solution}\label{sec:CdL}

Before we analyze Eq.~\eqref{eqn:general_eom} in the presence of black holes, it is quite important to review the dynamics in the absence of black holes, i.e. the CdL scenario~\cite{Coleman:1980aw}. Our goal is not to go over some known results but rather to stress the physical meaning of the solution and gain the understanding that will become essential later on when we include black holes. In Eq.~\eqref{eqn:general_eom} the metric function is that of Euclidean de-Sitter space written in static coordinates
\begin{align}
f_\pm (r_\pm) = 1- \frac{\Lambda_\pm r_\pm^2}{3} \ \ ,
\end{align}
and thus
\begin{align}\label{eq:cdleom}
\dot{R}^2 = 1 - \zeta R^2\, , \quad \zeta := 4\pi^2 G^2\sigma^2 + \frac{4\pi G(\epsilon_+ + \epsilon_-)}{3} + \frac{(\epsilon_+ - \epsilon_-)^2}{9\sigma^2}  \ \ .
\end{align}
The solution of the above equation is immediate, and one can then plug back in Eq.~\eqref{eqn:juncone} to find $\tau_\pm (\lambda)$,
\begin{align}\label{eq:wallcdl}
R(\lambda) = \frac{1}{\zeta} \cos(\zeta \lambda)\, , \quad \tau_\pm(\lambda) = l_\pm \tan^{-1} \left( \frac{\sin(\zeta \lambda)}{\sqrt{\zeta^2 l_\pm^2 -1}} \right)\, , \quad l_\pm^2 := \frac{3M_\textrm{P}^2}{\epsilon_\pm} \ \ ,
\end{align}
where the solution exists if and only if $\zeta^2 l_\pm^2 > 1$. Therefore, in static coordinates the CdL solution is oscillatory with period $\beta_\lambda = 2\pi/\zeta$. Based on that one might be tempted to think that the solution possesses an $O(3)$ symmetry, nevertheless, this is {\em not} true and the $O(4)$ symmetry of the CdL solution is nothing but hidden by the choice of coordinates. 

The symmetry can be made manifest in the global coordinates of Euclidean de-Sitter space, or more precisely angular coordinates on $S^4$. The argument goes as follows. The geometry of the thin-wall instanton is simple; we have two 4-dimensional spheres, with radii $l_+$ and $l_-$, which are glued at some polar angle that will be determined below. First, the coordinate transformation we need is of the form 
\begin{align}\label{eq:cordtrans}
r_\pm = l_\pm \sin(\xi_\pm) \sin\chi, \quad \tau_\pm = l_\pm \tan^{-1}\bigg[\tan(\xi_\pm) \cos\chi \bigg] \ \ ,
\end{align}
and leads to the following metric of Euclidean dS
\begin{align}\label{eq:4sphere}
ds_\pm^2 = l_\pm^2 \bigg( d\xi_\pm^2 + \sin^2(\xi_\pm) d\Omega_{(3)}^2  \bigg)\ \ ,
\end{align}
where $d\Omega_{(3)}^2$ is the metric on a unit 3-sphere. Eq.~\eqref{eq:4sphere} is nothing but the standard metric on a 4-sphere with radius $l_\pm$. These coordinates cover the whole sphere. To prove the $O(4)$ invariance of the solution, we just need to show that the wall motion, represented by Eq.~\eqref{eq:wallcdl}, becomes static in global coordinates and, in addition, find the polar angles where the two spheres are glued. From Eq.~\eqref{eq:wallcdl} we notice that
\begin{align}\label{eq:paramCDL}
\cos^2\left(\frac{\tau_\pm(\lambda)}{l_\pm} \right) \bigg(R^2(\lambda) + l_\pm^2 \tan^2\left(\frac{\tau_\pm(\lambda)}{l_\pm} \right) \bigg) = \frac{1}{\zeta^2} \ \ ,
\end{align}
and thus using Eq.~\eqref{eq:cordtrans} we find the wall to be located at
\begin{align}\label{eq:zetapm}
l_\pm \sin \xi^{\rm wall}_\pm = \frac{1}{\zeta} \ \ .
\end{align}
If we recall that $\zeta^2 l_\pm^2 > 1$, then the above equation possesses a solution. As we anticipated each $S^4$ is cut at the polar angle, $\xi$, determined by Eq.~\eqref{eq:zetapm} for each 4-sphere and then smoothly glued to form the static CdL instanton. The most important observation is that the solution manifestly displays an $O(4)$ symmetry, and therefore it represents a tunneling configuration that proceeds via vacuum decay.

\subsection{Including Black holes}

Using the formalism outlined above we include a static black hole in the spacetime and solve Eq.~\eqref{eqn:general_eom}. The metric is that of a static Schwarzschild de-Sitter black hole continued to Euclidean space. Hence, in Eq.~(\ref{eqn:general_eom}) we substitute
\begin{equation}\label{BHfofr}
f_\pm(r)=1-\frac{2GM_\pm}{r}-\frac{\Lambda_\pm r^2}{3} \ \ ,
\end{equation}
where $M$ is the ADM mass. As a reminder $+$ represents the false vacuum outside the bubble, while $-$ is the true vacuum inside the bubble. Therefore, $M_+$ is the seed black hole mass around which the bubble nucleates, $M_-$ is the remnant black hole mass and finally $\Lambda_{+(-)}$ is the cosmological constant in the false (true) vacuum. Throughout our analysis we will instead write the cosmological constants in terms of vacuum energies, $\epsilon_\pm=M_\textrm{P}^2\Lambda_\pm$, allowing a more transparent comparison to quantities from the scalar field theory. Explicitly, in our case we have
\begin{align}\label{vaceps}
\Lambda_+ = \Lambda_0, \quad \Lambda_- = \Lambda_0 - \frac{\epsilon_{\varphi}}{M_\textrm{P}^2}  \ \ ,
\end{align}
where $\epsilon_{\varphi}$ is the vacuum energy density of the true vacuum. In addition, we have the Hawking temperatures for both the black hole and cosmological horizons
\begin{align}\label{Hawtemps}
\beta_h = \frac{4\pi}{|f^\prime(r_h)|}, \quad \beta_{\rm cos} = \frac{4\pi}{|f^\prime(r_c)|} \ \ ,
\end{align}
where $r_h$ and $r_c$ are the two positive roots of metric function. To ensure the presence of a horizon, or rather the absence of a naked singularity, we have the following inequality
\begin{align}\label{nonext}
\Lambda_\pm < \frac{1}{(3GM_\pm)^2} \ \ .
\end{align}

Before solving the system, we pause to comment on the validity of the thin-wall approximation, which in standard QFT just amounts to having the radius of the bubble much larger than its thickness. The thickness of the bubble, in the thin-wall approximation, is roughly given by the inverse mass of the field, while the radius of the bubble is inversely proportional to the energy density difference. Therefore, for small energy density the thin-wall approximation is valid. We want know if the presence of gravity requires any new conditions so as to validate the thin-wall approximation. To this aim, let us substitute the metric function, Eq.~\eqref{BHfofr}, into the equation of motion for the scalar field, Eq.~\eqref{eombasic}, and find
\begin{align}\label{eq:eomphi}
\left( f(r) \, \partial_r^2 + f^\prime(r) \partial_r + \frac{2f(r)}{r} \partial_r \right) \varphi(r) = \frac{dV\left(\varphi,T\right)}{d\varphi} \ \ ,
\end{align}
where, for simplicity, we focus on a static configuration. Operationally, the thin-wall approximation allows us to drop terms that go as $1/r$ in the equation of motion, which are negligible near the bubble wall. Inspection of Eq.~\eqref{eq:eomphi} shows that the presence of the metric function does not change anything, in particular, the term with $f^\prime(r)$ scales as $f(r)/r$ and thus can be ignored for large bubbles. Therefore, even in our case, the thin-wall approximation remains valid as long as the radius of the bubble is much larger than the wall thickness.

\begin{figure}[t]
	\center
  	\includegraphics[width=0.65\textwidth]{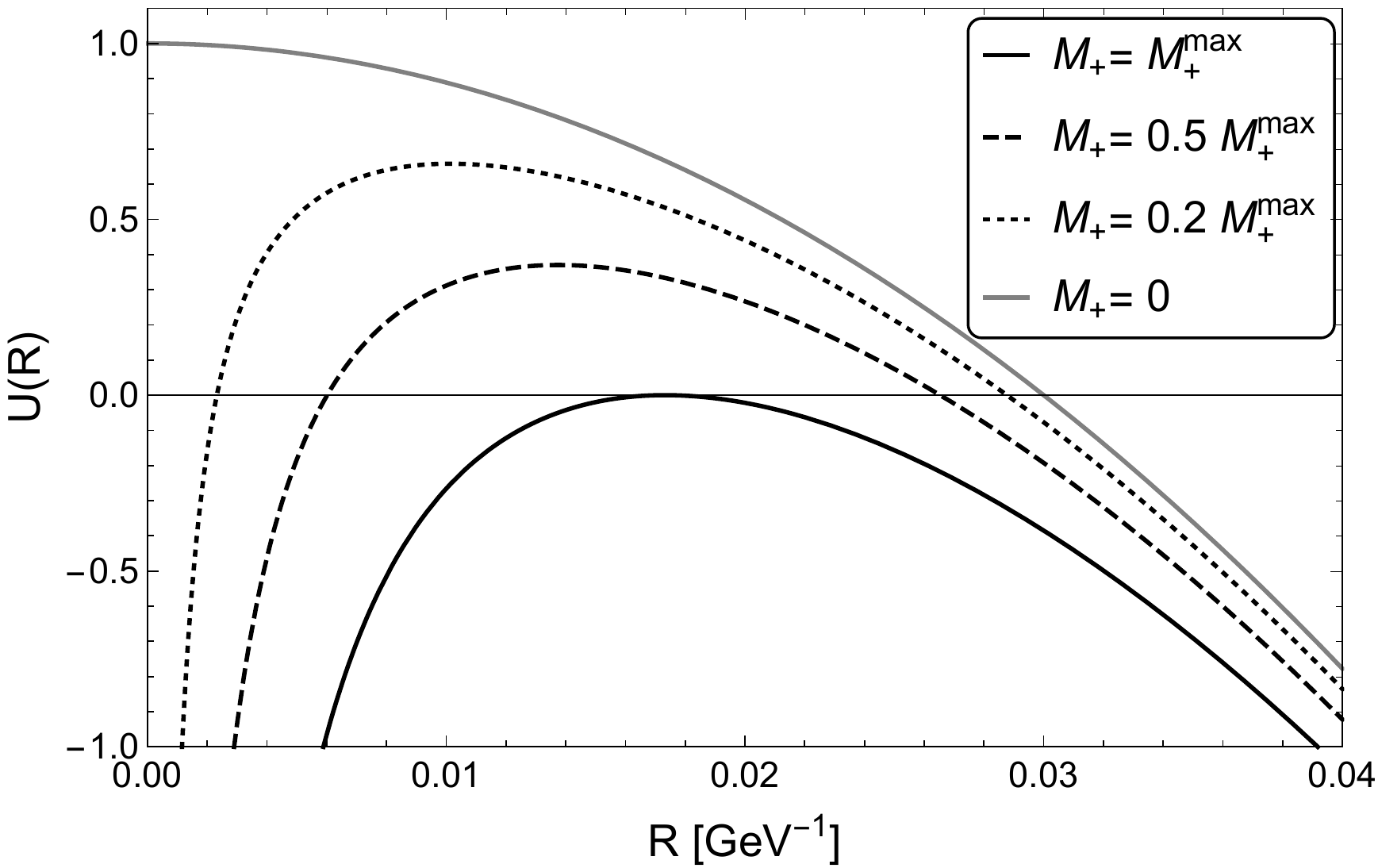}
  	\caption{The potential $U(R)$ of Eq.~\eqref{eqn:BH_eom} describing the bubble wall trajectory for varying seed black hole masses $M_+$ with $M_-=M_+$ and example parameter values $\sigma=10^4\, \text{GeV}^3$, $\epsilon_\varphi=10^6\, \text{GeV}^4$ and $\epsilon_+-\epsilon_\varphi\sim\epsilon_\textrm{DE}$ where $\epsilon_\textrm{DE}$ is the dark energy scale (setting $\epsilon_+-\epsilon_\varphi=0$ would not noticeably change the results). With these parameters, $M_+^\textrm{max}=3.5\times 10^{17}\, M_\textrm{P}$.}
  	\label{fig:U_pot}
\end{figure}

\begin{figure}[t]
	\center
  	\includegraphics[width=0.65\textwidth]{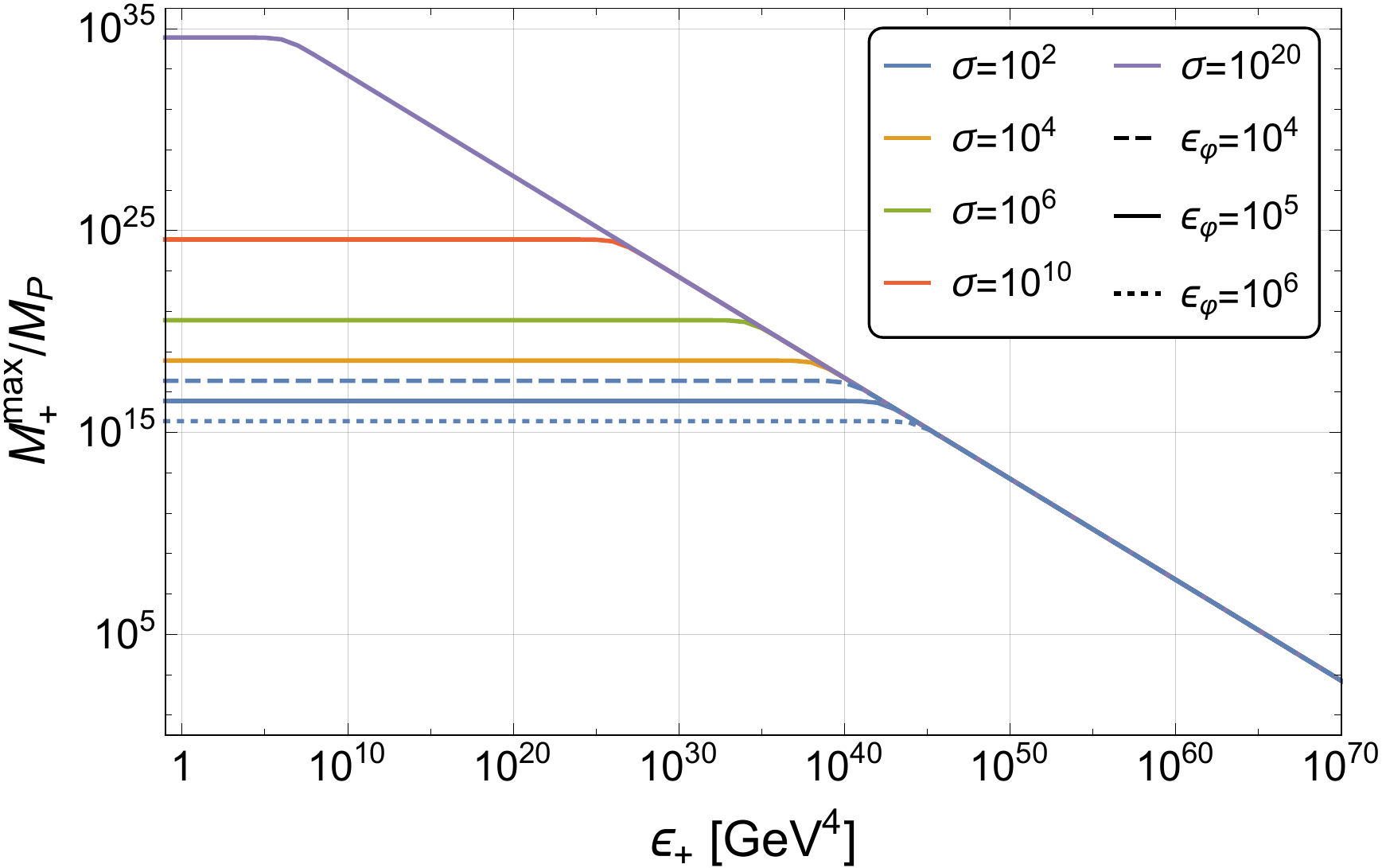}
  	\caption{The absolute maximum seed black hole mass $M_+^\text{max}$ as a function of the false vacuum energy $\epsilon_+$ for varying surface tension $\sigma$ (in $\text{GeV}^3$) and change in vacuum energy $\epsilon_\varphi$ (in $\text{GeV}^4$) values. Colour represents a changing $\sigma$ value. Dashed, solid and dotted lines represent increases in $\epsilon_\varphi$ respectively. At large $\epsilon_+$ values the dependence on $\sigma$ and $\epsilon_\varphi$ is alleviated. Notice how, at lower $\epsilon_+$, it is the ratio $\sigma/\epsilon_\varphi$ that determines $M_+^\textrm{max}$ and not their separate values; for example $\sigma=10^4,\, \epsilon_\varphi=10^5$ and $\sigma=10^{10},\, \epsilon_\varphi=10^{11}$ would have the same $M_+^\textrm{max}$ value.}
  	\label{fig:M2Crit_vs_e2}
\end{figure}

We finally substitute the metric function into Eq.~\eqref{eqn:general_eom} to find
\begin{alignat}{2}\label{eqn:BH_eom}
\dot{R}^2 & = 1 &&- R^2\left[ 4\pi^2 G^2\sigma^2 + \frac{4\pi G(2\epsilon_+ - \epsilon_\varphi)}{3} + \frac{\epsilon_\varphi^2}{9\sigma^2} \right]\nonumber \\
        & &&- \frac{1}{R}\left[ G(M_++M_-) + \frac{(M_+-M_-)\epsilon_\varphi}{6\pi\sigma^2} \right]\nonumber \\
        & &&- \frac{1}{R^4}\left[ \frac{(M_+-M_-)^2}{16\pi^2\sigma^2} \right]\nonumber \\
        & := U&&(R) \ \ .
\end{alignat}
Two parameters from the scalar field theory, the surface tension $\sigma$ and the vacuum energy $\epsilon_\varphi$, now appear in the equation of motion of the wall. For cosmological phase transitions the origin of the seeds will be primordial black holes. Solutions of Eq.~(\ref{eqn:BH_eom}) clearly depend on all the parameters $M_\pm$, $\epsilon_+$, $\sigma$ and $\epsilon_\varphi$. Thus, for starters, it is crucial to know if there exists any {\em absolute} bound on any single parameter. Fig.~\ref{fig:U_pot} shows the generic behavior of the potential, $U(R)$, for an arbitrary choice of parameters. We immediately observe that for Eq.~(\ref{eqn:BH_eom}) to possess a solution the potential must be non-negative over some portion of its domain.
First, let us restrict the mass of the black hole post-transition to be smaller or equal to that of the seed, $M_-\leq M_+$. Although in principle it could be larger, restricting $M_-$ suffices for our purposes because, as we will show, the dominant tunneling configuration turns out to have the minimum accessible $M_-$ at fixed seed $M_+$. Moving on, there exists an absolute bound on the seed mass as a function of the other variables, i.e $(\epsilon_+, \epsilon_\varphi, \sigma)$. It is straightforward to find this absolute value on $M_+$, at least numerically, by studying the single extremal point of the potential, i.e. 
\begin{align}\label{eqn:R_s}
U^\prime(R_{\rm ext}) = 0 \ \  ,
\end{align}
 where all variables are fixed. Here, $R_{\rm ext}$ denotes the radius of the bubble wall at the extremal point of the potential. No solutions exist for Eq.~(\ref{eqn:BH_eom}) if and only if the value of the potential, at its critical point, is negative
\begin{align}\label{eqn:sol_cond}
U(R_{\rm ext}) < 0  \ \ .
\end{align}
A close look at the potential illustrates the roles of $M_\pm$; increasing $M_+$ lowers the potential while increasing $M_-$ raises the potential. Therefore, the maximum value of $U(R_{\rm ext})$ is obtained by setting $M_-$ to its maximal value, i.e. $M_-=M_+$. Hence, using this in combination with the condition of Eq.~\eqref{eqn:sol_cond} gives us an absolute upper bound on the seed mass, $M_+^\text{max}$, that we determine numerically in Fig.~\ref{fig:M2Crit_vs_e2}. Essentially, $M_+^\text{max}$ remains constant up to a certain value of $\epsilon_+$, which depends on $\sigma$ and $\epsilon_\varphi$, after which it decreases dramatically. Larger $\sigma$ and smaller $\epsilon_\varphi$ reduce the plateau portion of $M_+^\text{max}$. It is also important to notice that all the curves in Fig.~\ref{fig:M2Crit_vs_e2} merge at high $\epsilon_+$ values, practically eliminating the dependence of $M_+^\text{max}$ on $\sigma$ and $\epsilon_\varphi$. 
Apart from $M_+$, all other variables are not constrained except by the phenomenology of the underlying physics model. Inspection of exact numbers shown in Fig.~\ref{fig:M2Crit_vs_e2} implies the non-trivial constraint imposed by the physics model, through $\sigma$ and $\epsilon_\varphi$, on the potential relevance of primordial black holes in the phase transition process. For example, for electroweak like phase transitions, we have the typical values
\begin{align}\label{devalues}
\sigma^{\text{EW}} \sim 10^4\, \rm GeV^3, \quad\epsilon^{\text{EW}}_\varphi \sim 10^7\, \rm GeV^4 \ \ ,
\end{align}
leaving the upper limit on $M_+^{\rm max}$  to be around $10^{16} M_{\rm P}$ (blue curve in Fig.~\ref{fig:M2Crit_vs_e2}). This shows that bubbles can only nucleate around relatively small black holes, roughly on the order of  $10^{-22}$ solar masses $(\sim 10^8\, \textrm{kg})$.

\subsection{Static bubbles \& periodic bounces}\label{conds}

Before we move to construct explicit solutions of Eq.~\eqref{eqn:BH_eom}, it is imperative to pause and understand their qualtitative nature and physical meaning. In the presence of black holes, it is far from trivial to directly interpret these thin-wall instantons, and uncover the role they play in the tunneling process. To this aim, we advocate a conservative approach and rely on asserting direct correspondence with standard tunneling configurations in QFT (see Sect.~\ref{sec:QFT}). Generically, Eq.~\eqref{eqn:BH_eom} displays two {\em disconnected} classes of solutions and we discuss each in turn.

\paragraph{The static branch:}The first class contains static solutions, i.e. $R(\lambda) = R_{\rm s}$. For each input value of the seed mass, $M_+$, there exists a unique static branch where the potential vanishes at its extremal point
\begin{align}\label{eqn:static_cond}
U(R_{\rm s}) = U^\prime(R_{\rm s}) = 0 \Rightarrow \dot{R}=0 \ \ .
\end{align}
This situation is depicted by the solid black curve in Fig.~\ref{fig:U_pot}. We first solve Eq.~\eqref{eqn:R_s} for $R_{\rm s}$, which makes the latter an explicit function of $M_-$, with all other parameters fixed. We then use Eq.~\eqref{eqn:static_cond} to determine the mass of the remnant black hole, that we denote by $M_-^{\rm s}$. With $M_-^{\rm s}$ in hand it is then straightforward to substitute back in and determine $R_{\rm s}$.

What is the physical significance of the static branch? Clearly, this class possesses a manifest $O(3)$ spherical symmetry and is independent of Euclidean time. Using the dictionary of QFT (see Sect.~\ref{sec:QFT}) these solutions are the equivalent of the typical $O(3)$ invariant {\em bubbles} familiar from finite-temperature phase transitions. In particular, and as we explained in Sect.~\ref{sec:QFT}, the transition in this case proceeds by thermal excitation over the potential barrier.

\paragraph{The oscillating branch:} The second class of solutions emerge when $M_- \neq M_-^{\rm s}$. For each value in the admissible range, i.e.
\begin{equation}\label{eqn:M1_range}
     M_- \in (M_-^{\rm s},M_+^{\rm} ] \ \ ,
\end{equation}  
the potential is positive semi-definite between $R_\text{min}$ and $R_\text{max}$, which satisfy
\begin{equation}
    U(R_\text{min})=U(R_\text{max})=0 \ \ .
\end{equation}
Between these two points, the wall separating the two vacua will {\em oscillate} indefinitely in Euclidean time, with a finite period $\beta_\lambda$. This period, crucially, is fully dictated by the dynamics of Eq.~\eqref{eqn:BH_eom} and is an explicit function of the input parameters in the theory. Therefore, Eq.~\eqref{eqn:BH_eom} possesses an infinite set of connected solutions parametrized by $M_-$ in the admissible range. Moreover, we also stress that, by virtue of Eq.~\eqref{eqn:juncone}, the period of Euclidean times in both vacua, i.e. $(\beta_{\tau_+},\beta_{\tau_-})$, are dictated dynamically.

We now ask the question: which, out of this infinite set, correspond to a valid tunneling configuration? Let us recall that the solution has an enhanced $O(4)$ symmetry {\em only} in the CdL case, $M_+ = M_- = 0$, as we have demonstrated in Sect.~\ref{sec:CdL}. Based on our discussion in Sect.~\ref{sec:QFT}, we interpret these finite-period oscillating solutions as the equivalent of the {\em finite-period bounces} which describe thermally-assisted tunneling in QFT. Therefore, for each input value of $M_+$ we demand the period of oscillation to match the inverse temperature of the system, i.e.
\begin{equation}\label{eqn:temp_cond}
    \beta_\lambda=\frac{1}{T} \ \ ,
\end{equation}
which presents a sufficient condition to single out a unique value for the remnant black hole mass, that we denote by $M_-^{\beta}$.

\paragraph{Quantitative analysis:} We are now rightly oriented to numerically construct the solutions we are interested in. The size of $\epsilon_\varphi$ is born out of the underlying theory, however, $\epsilon_+$ may be thought of as a totally free parameter. We make two choices for $\epsilon_+$. First, in order to make contact with cosmology, $\epsilon_+$ is fixed such that the vacuum energy after the transition matches the dark energy scale. Second, we pick a huge value of $10^{40}\, \text{GeV}^4$ to suppress the dependence of our analysis on the two other parameters, $\epsilon_{\varphi}$ and $\sigma$, as suggested by Fig.~\ref{fig:M2Crit_vs_e2}. 

\begin{figure}[t]
     \centering
     \begin{subfigure}{0.49\textwidth}
     \includegraphics[scale=0.50]{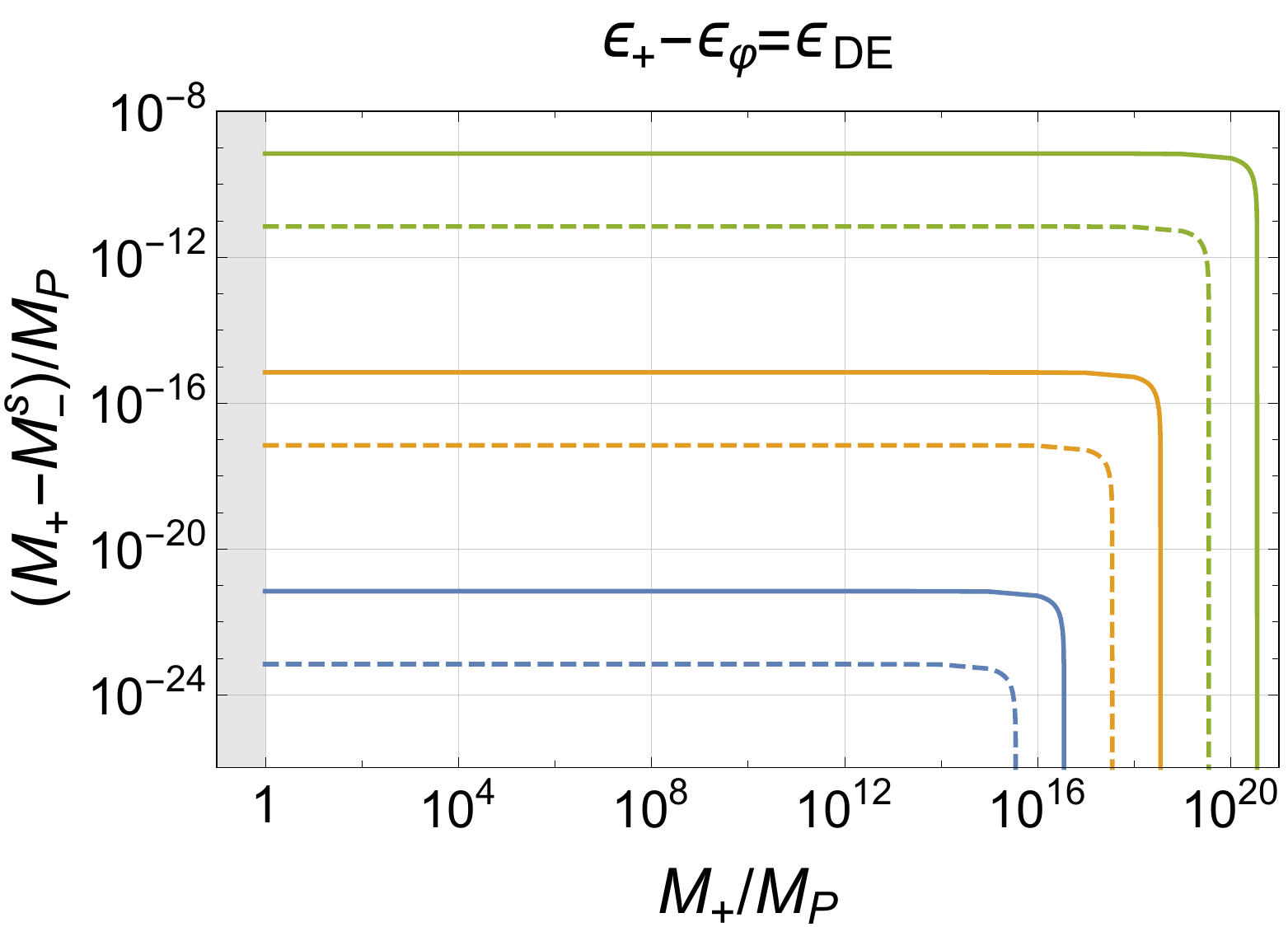}
     \end{subfigure}
     \hfill
     \begin{subfigure}{0.49\textwidth}
     \includegraphics[scale=0.50]{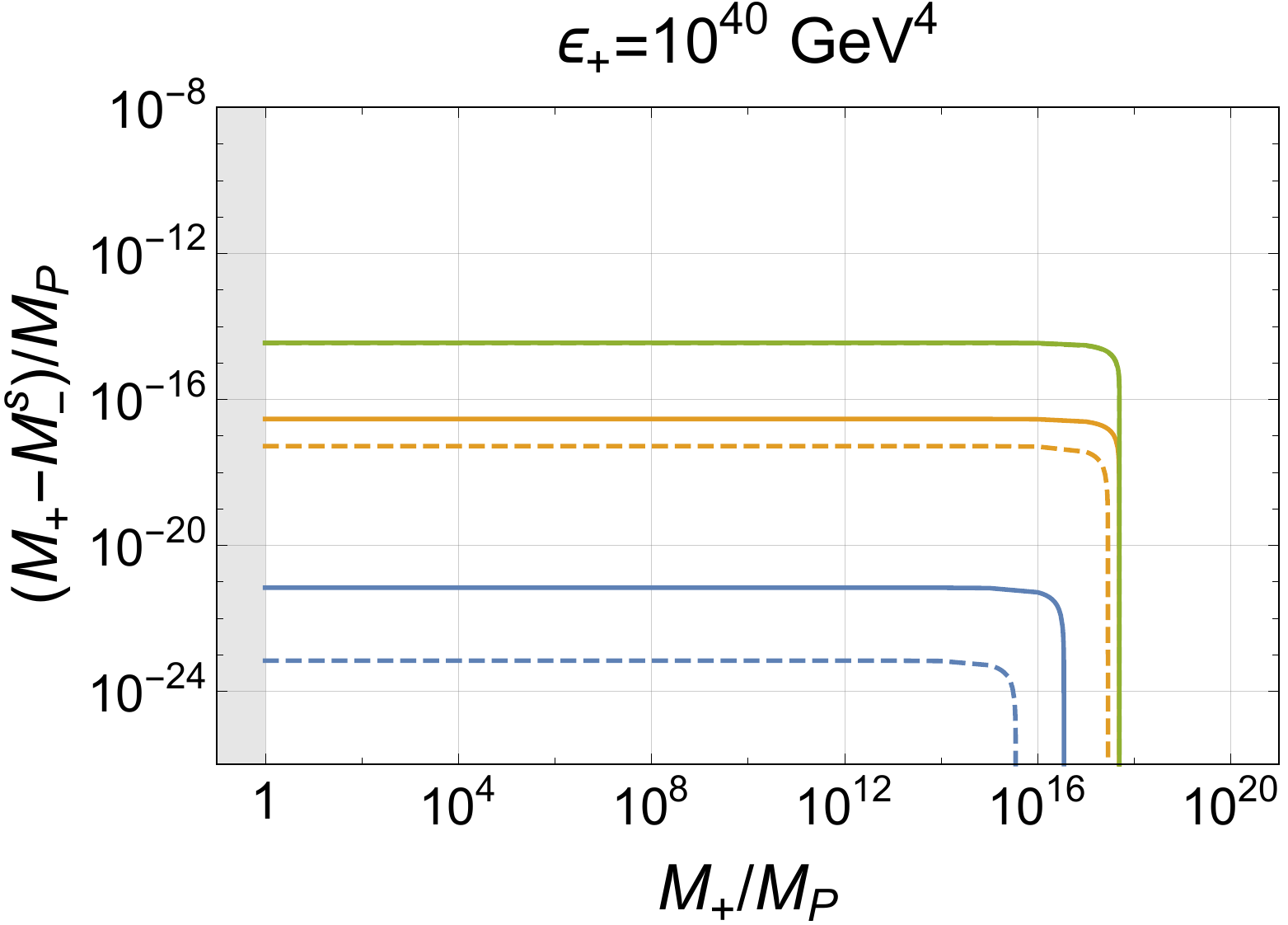}
     \end{subfigure}
     \vfill
     \begin{subfigure}{0.49\textwidth}
     \centering
     \includegraphics[scale=0.65]{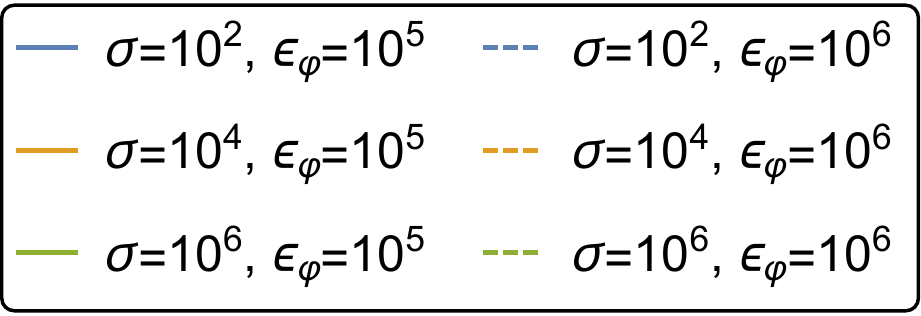}
     \end{subfigure}
     \caption{$M_+-M_-^s$ as a function of $M_+$ for various $\sigma$ (in $\textrm{GeV}^3$) and $\epsilon_\varphi$ (in $\textrm{GeV}^4$) values. The grey shaded region is sub-Planckian mass, $M_+<M_\textrm{P}$. Left: $\epsilon_+-\epsilon_\varphi=\epsilon_\textrm{DE}$. Right: $\epsilon_+=10^{40}\, \text{GeV}^4$.}
     \label{fig:M1_vs_M2}
\end{figure}

\begin{figure}[t]
     \centering
     \begin{subfigure}{0.49\textwidth}
     \includegraphics[scale=0.50]{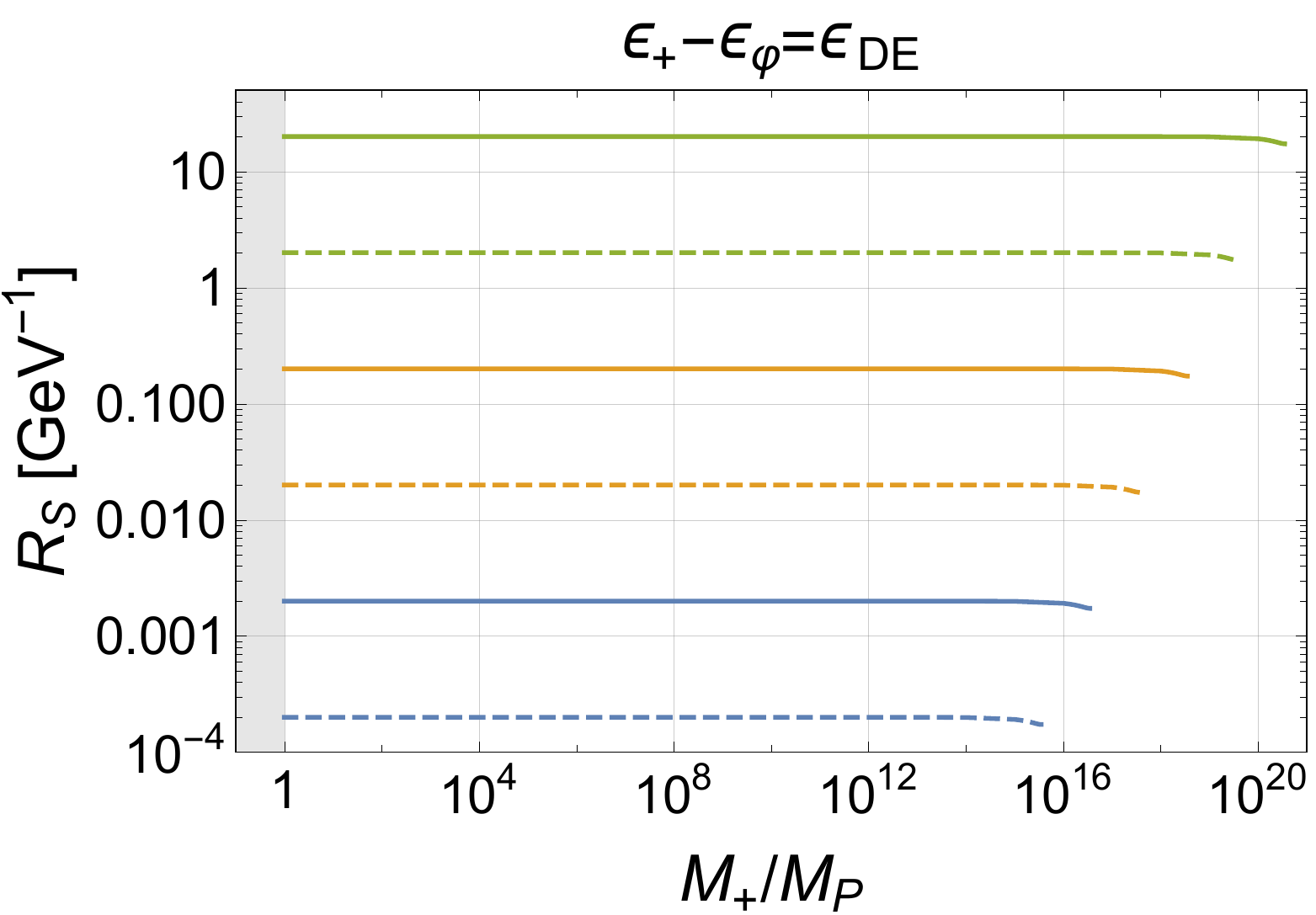}
     \end{subfigure}
     \hfill
     \begin{subfigure}{0.49\textwidth}
     \includegraphics[scale=0.50]{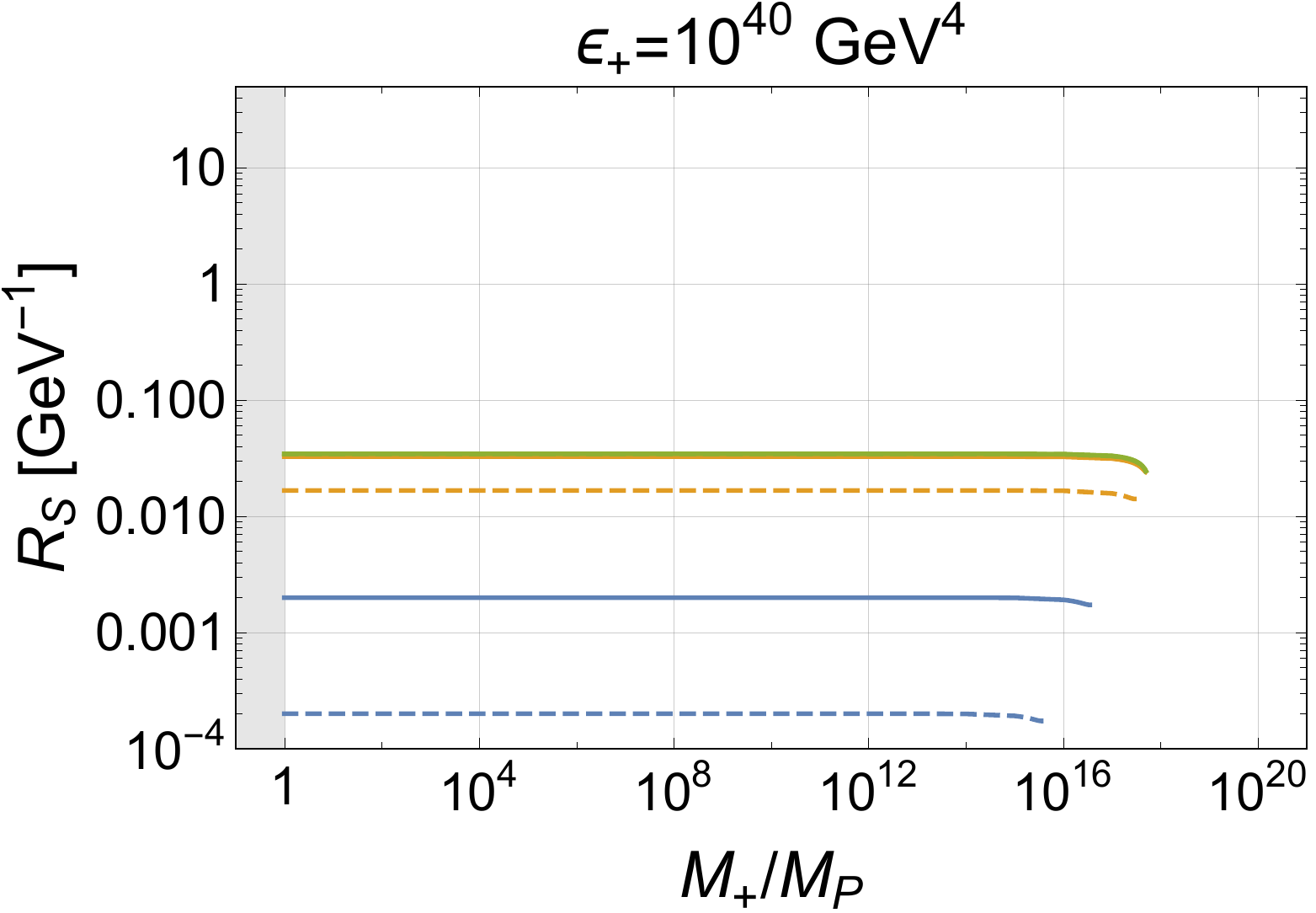}
     \end{subfigure}
     \vfill
     \begin{subfigure}{1\textwidth}
     \centering
     \includegraphics[scale=0.65]{Figures/plot_legend.pdf}
     \end{subfigure}     
     \caption{$R_s$ as a function of $M_+$ for various $\sigma$ (in $\textrm{GeV}^3$) and $\epsilon_\varphi$ (in $\textrm{GeV}^4$) values. The grey shaded region is sub-Planckian mass, $M_+<M_\textrm{P}$. Left: $\epsilon_+-\epsilon_\varphi=\epsilon_\textrm{DE}$. Right: $\epsilon_+=10^{40}\, \text{GeV}^4$.}
     \label{fig:Rs_vs_M2}
\end{figure}

We start by analyzing the static branch. Given an admissible value for $M_+$, fixing $(\epsilon_+, \epsilon_\varphi, \sigma)$, we solve Eq.~\eqref{eqn:static_cond} for $M_-$. This particular value, that we denote by $M_-^s$, provides a lower bound on the mass of the remnant black hole. In Fig.~\ref{fig:M1_vs_M2} we provide the values of $M_-^s$ for various parameter choices. We observe three striking features. First, $M_-^s$ is generically very close to the seed mass $M_+$. Second, the difference $M_+-M_-^s$ remains essentially constant for almost the whole admissible range of $M_+$. Third, increasing $\sigma$ substantially increases the difference while increasing $\epsilon_\varphi$ decreases the difference. Note that the endpoint of each line corresponds to the configuration where $M_+=M_+^\text{max}$. Moving on, in Fig.~\ref{fig:Rs_vs_M2} we show the size of the static bubble, $R_s$, for the same parameter choices. We observe two main features. First, the seed black hole mass $M_+$ has an {\em insignificant} effect on $R_{\rm s}$ compared to $\sigma$ and $\epsilon_\varphi$. Second, increasing $\sigma$ by an order of magnitude noticeably increases the size of the bubble, while increasing $\epsilon_\varphi$ has an equal but opposite effect.


\begin{figure}[ht]
    \centering
    \includegraphics[width=0.50\textwidth]{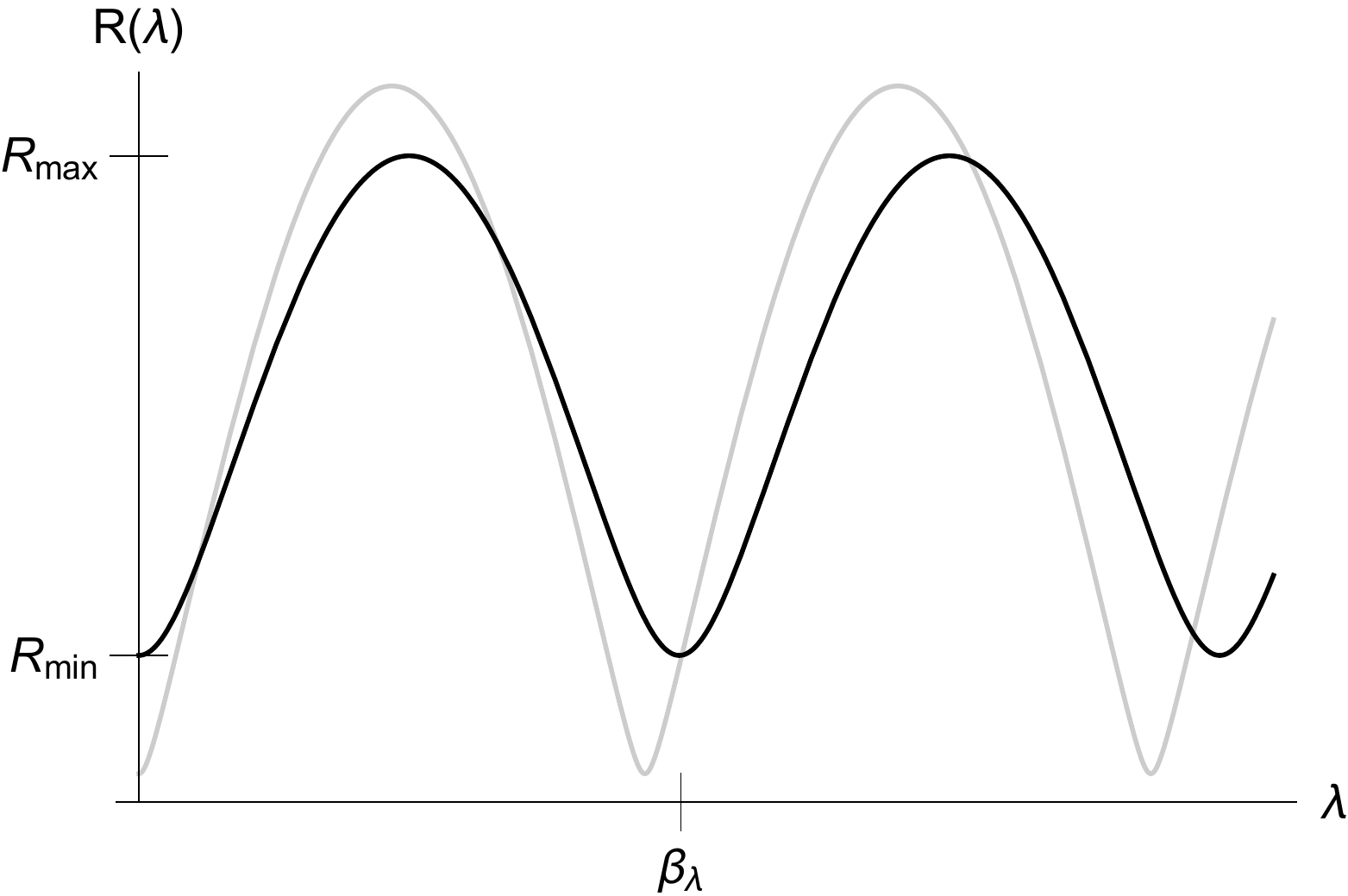}
    \caption{A sketch of two oscillating solutions. The labels correspond to the black line, which has a period $\beta_\lambda$. The minimum $R_\text{min}$ and maximum $R_\text{max}$ are the roots of the potential $U(R)$ given in Eq.~\eqref{eqn:BH_eom}.}
    \label{fig:osc_sol}
\end{figure}

\begin{figure}[ht]
     \centering
     \begin{subfigure}{0.49\textwidth}
     \includegraphics[scale=0.50]{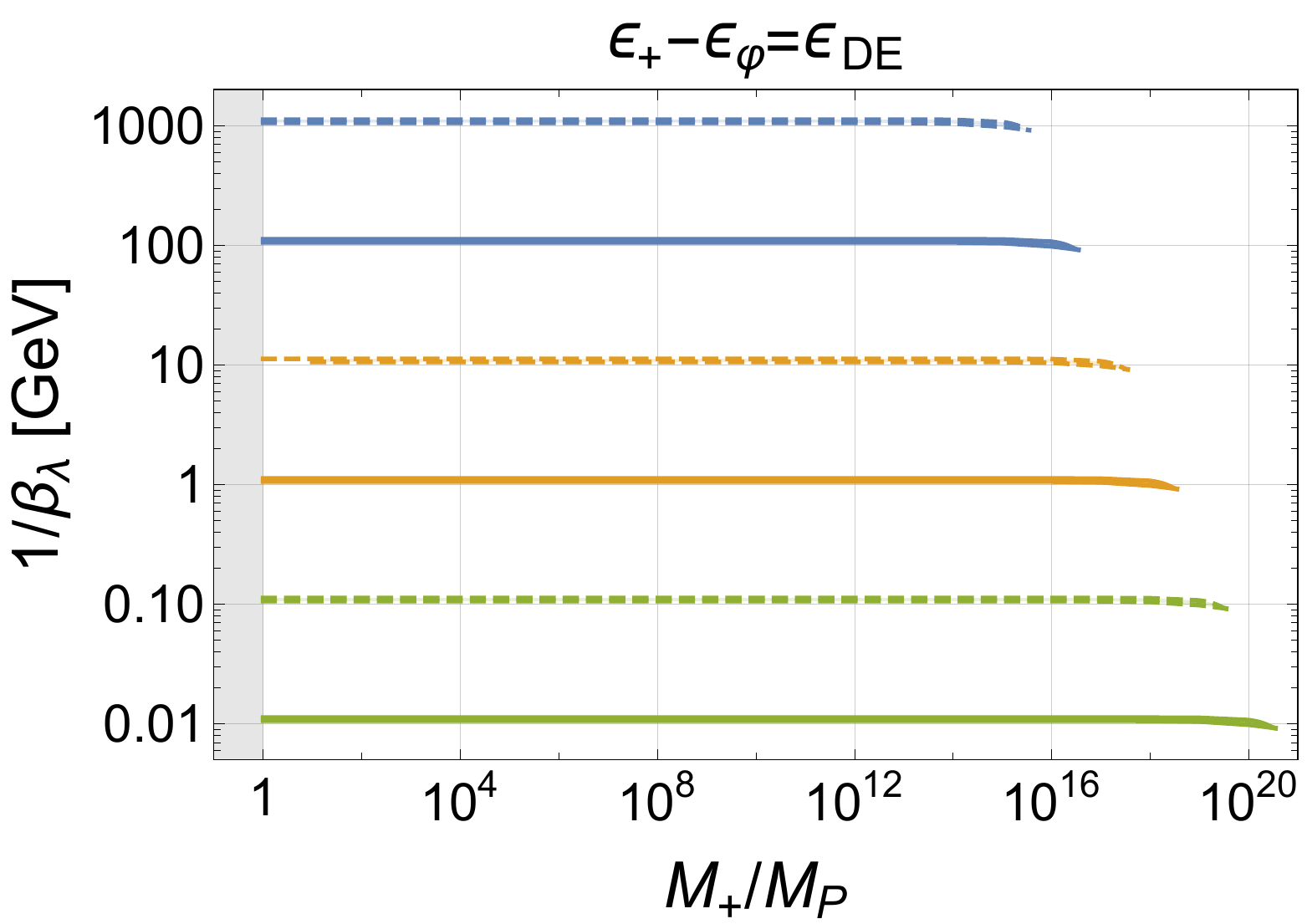}
     \end{subfigure}
     \hfill
     \begin{subfigure}{0.49\textwidth}
     \includegraphics[scale=0.50]{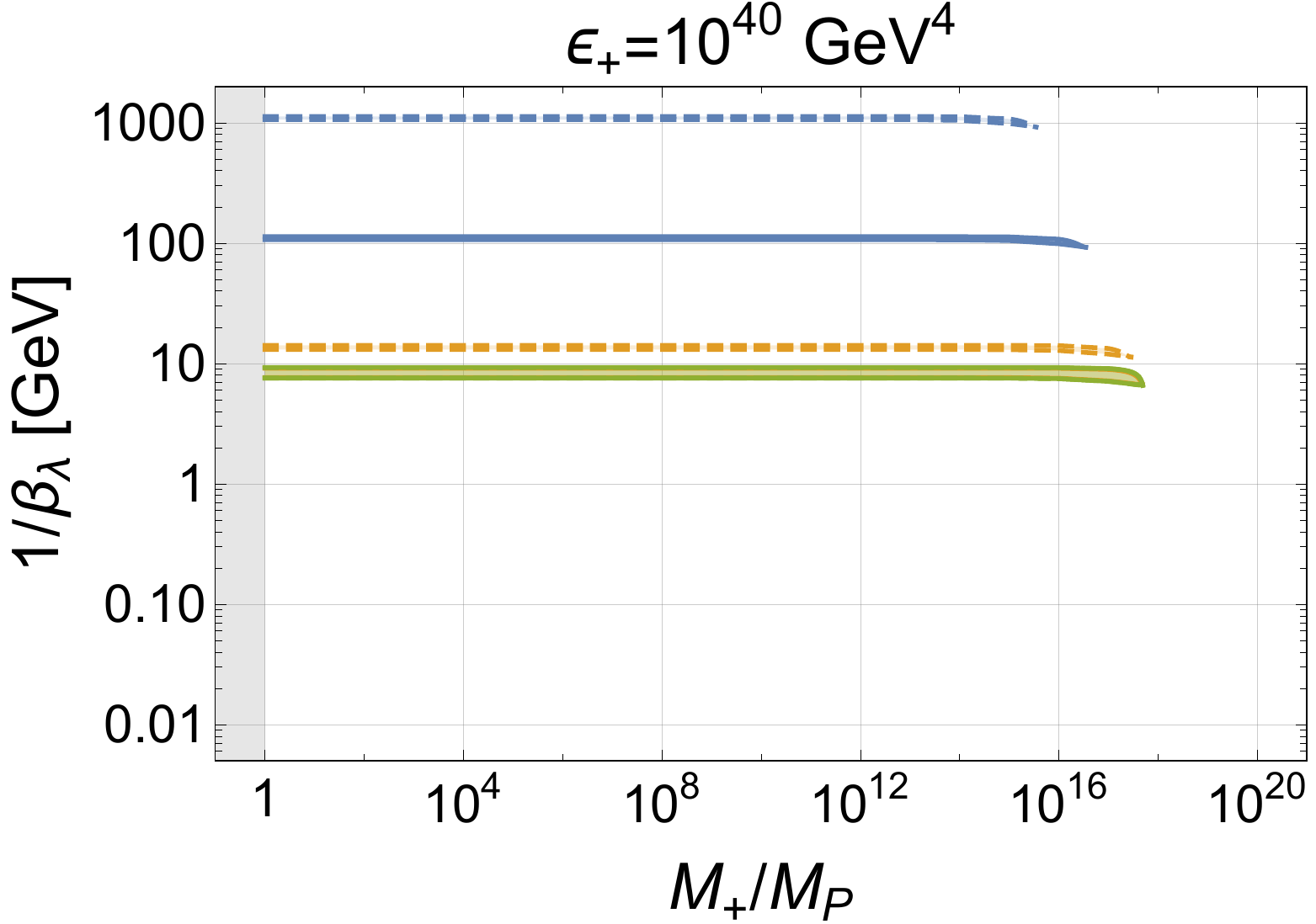}
     \end{subfigure}
     \vfill
     \begin{subfigure}{0.49\textwidth}
     \includegraphics[scale=0.50]{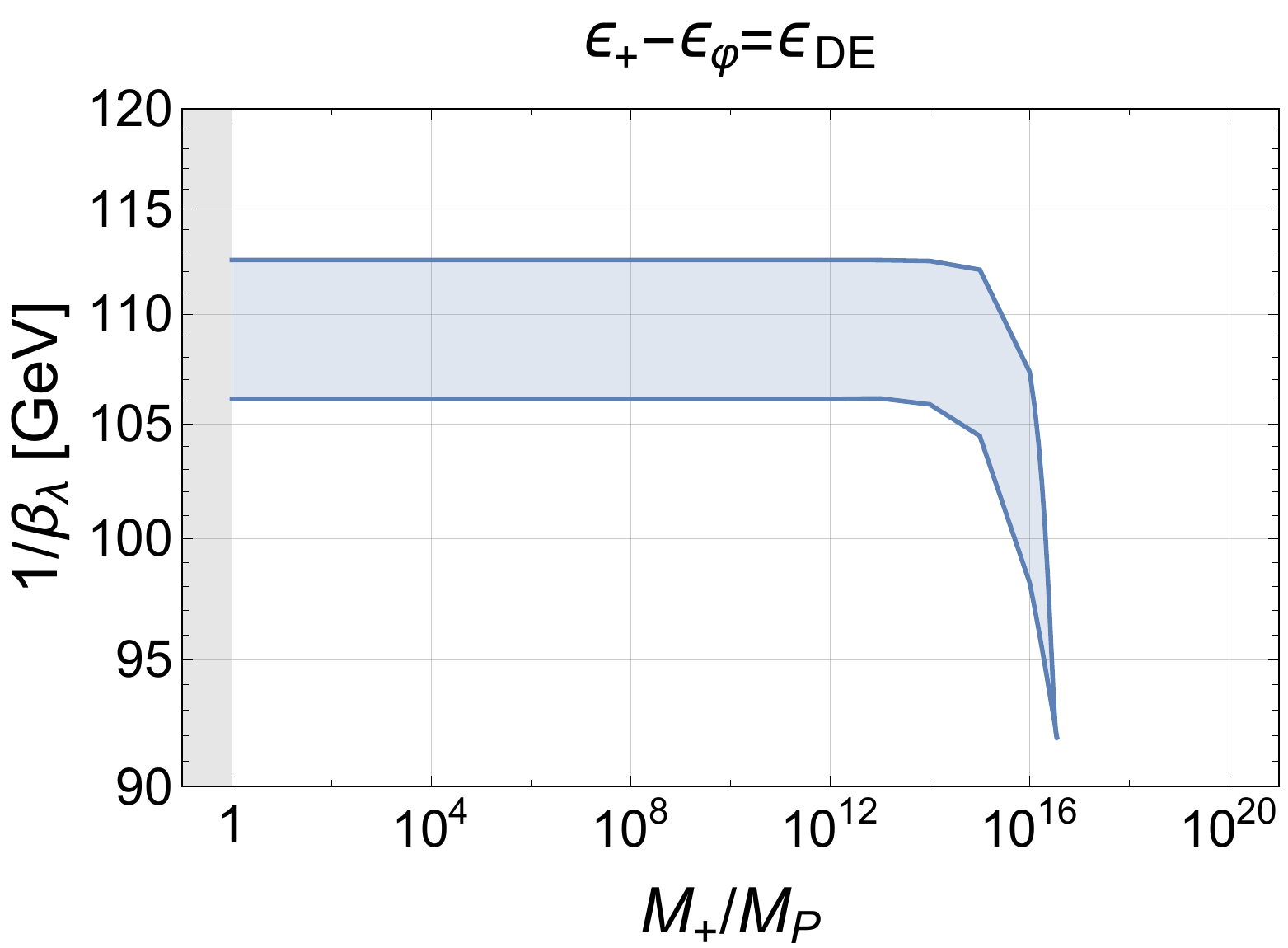}
     \end{subfigure}
     \hfill
     \begin{subfigure}{0.49\textwidth}
     \centering
     \includegraphics[scale=0.65]{Figures/plot_legend.pdf}
     \end{subfigure}
     \caption{The inverse period $1/\beta_\lambda$ of the oscillating solutions as a function of $M_+$ for various $\sigma$ (in $\textrm{GeV}^3$) and $\epsilon_\varphi$ (in $\textrm{GeV}^4$) values. The grey shaded region is sub-Planckian mass, $M_+<M_\textrm{P}$. Top Left: $\epsilon_+-\epsilon_\varphi=\epsilon_\textrm{DE}$. Top Right: $\epsilon_+=10^{40}\, \text{GeV}^4$. Bottom Left: Focus on the $1/\beta_\lambda$ region for $\sigma=10^2\ \text{GeV}^3$, $\epsilon_\varphi=10^5\ \text{GeV}^4$ and $\epsilon_+-\epsilon_\varphi=\epsilon_\textrm{DE}$. For the $1/\beta_\lambda$ regions, the upper limit is given by approaching the static limit $M_-\rightarrow M_-^s$, while the lower limit is given by $M_-=M_+$.}
     \label{fig:T_vs_M2}
\end{figure}

The second class of solutions are those which are periodic. In order to show the meaning of the quantities $R_\text{min}$, $R_\text{max}$ and $\beta_\lambda$, a sketch is shown in Fig.~\ref{fig:osc_sol}.
The valid solution describing a finite-period bounce satisfies Eq.~\eqref{eqn:temp_cond}; therefore, it is essential to understand the dependence of the period $\beta_\lambda$ on the input parameters.
Fig.~\ref{fig:T_vs_M2} plots the inverse period, $1/\beta_\lambda$, as a function of $M_+$ for various $\sigma$ and $\epsilon_\varphi$ values.
It is important to note that $1/\beta_\lambda$ has a mild dependence on $M_-$ as demonstrated in the bottom left plot of Fig.~\ref{fig:T_vs_M2}. We observe the following features. First, $1/\beta_\lambda$ remains essentially constant with respect to $M_+$ until it approaches $M_+^\text{max}$, at which point it sharply decreases to a finite minimum value. However, changing $\sigma$ and $\epsilon_\varphi$ drastically alter $1/\beta_\lambda$ in comparison to $M_+$. Second, the inverse period increases by an order of magnitude if either $\sigma$ decreases or $\epsilon_\varphi$ increases by an order of magnitude. This essentially means that the characteristics of the particle physics model controls whether or not we have a valid oscillating tunneling configuration. 
Given a set of $(\epsilon_+,\epsilon_\varphi,\sigma,M_+)$, the finite-period bounce satisfying the matching condition Eq.~\eqref{eqn:temp_cond} is obtained by continuously varying $M_-$ until $M_-^\beta$ is found. If no solution is found then black holes have no effect on tunneling via finite-period bounces and we are only left with the static branch.

Finally, we would like to highlight the importance of $M_+^\textrm{max}$. Consider the $1/\beta_\lambda$ plots of Fig.~\ref{fig:T_vs_M2}.  As we alluded to before, a very large value of $\epsilon_+$ constrains the limiting values of $M_+^{\rm max}$ to be identical irrespective of the other parameters. That is the reason behind the difference in the top two plots of Fig.~\ref{fig:T_vs_M2}, where we see the merger of many of the lines as we move from the left to the right plot. The message is clear and interesting: it is not the independent values of $\sigma$, $\epsilon_\varphi$ and $\epsilon_+$ that control the period of oscillation, but rather it is the \emph{combination} of the three parameters that sets the actual value of $M_+^\text{max}$, as shown in Fig.~\ref{fig:M2Crit_vs_e2}.

\paragraph{Qualitative final comments:} Having explicitly described the solutions to Eq.~\eqref{eqn:BH_eom} and their physical interpretations, one cannot help but inquire whether the dynamics of Eq.~\eqref{eqn:BH_eom} admit a solution that we could interpret as a zero-temperature bounce. In standard QFT, the special feature of vacuum tunneling manifests in the $O(4)$ symmetry of the bounce. To answer our question, therefore, one needs to inspect if a particular solution to Eq.~\eqref{eqn:BH_eom} could possess a {\em hidden} $O(4)$ symmetry.

Here we can make the analogy with the isolated case $M_-=M_+=0$ (CdL) that was discussed in Sect.~\ref{sec:CdL}. There, the specific sinusoidal form of the solution, Eq.~\eqref{eq:wallcdl}, was in fact hiding the symmetry, which becomes manifest when the solution is written in global coordinates. Now in the presence of black holes, the solution $R(\lambda)$ plotted in Fig.~\ref{fig:osc_sol} shows that the existence of such a hidden symmetry is no longer possible. Simply put, any solution to Eq.~\eqref{eqn:BH_eom} can generally be written as a Fourier sum, with frequencies $\omega_n  = 2\pi n /\beta_\lambda$ where $n$ is an integer. The existence of an infinite tower of Fourier modes confirms that a hidden $O(4)$ symmetry is not possible.  In essence, the wall, in the presence of black holes, is not following a simple parametric equation, in contrast to Eq.~\eqref{eq:paramCDL}.

\section{The Tunneling Exponent}\label{sec:bounce_action}

We now move to evaluate the Euclidean action given the thin-wall solutions constructed in the last section. This is a standard computation, and the only subtlety concerns the presence of conical sections in the geometry. The latter arise due to the mismatch between the oscillation period, $\beta_\lambda$, and the Hawking temperatures of both horizons. In the case of the Einstein-Hilbert action, there is a systematic technique to compute the contribution of the conical singularity to the action. In summary, consider a 4-dimensional spacetime which contains one or more conical deficit angles, $\alpha_i$, the integral over the Ricci scalar curvature reads~\cite{Fursaev:1995ef}
\begin{equation}
\int d^4x\, \sqrt{g}\, \mathcal{R} = \sum_{i} 4\pi (1 - \alpha_i ) A_i + \int d^4x\, \sqrt{g}\,  \mathcal{R}_{\rm reg} \ \ ,
\end{equation}
where $\mathcal{R}_{\rm reg}$ is the non-singular part of the Ricci scalar and $A_i$ is the 2-dimensional area of the conical surface. In the thin-wall approximation, the bubble action is composed of two components. First, we have the {\em bulk} component represented by contributions of both the false and true vacua. The second component is a {\em surface} contribution represented by the bubble wall. 

The bulk Euclidean action is very simple and reads
\begin{align}\label{bulkaction}
I_E^{\text{bulk}}(g,\varphi) = - \frac{M_\textrm{P}^2}{2} \int d^4x\, \sqrt{g}\, \left(\mathcal{R} - 2 \Lambda \right) \ \ ,
\end{align}
where the cosmological constant contains the contribution of the potential energy density of the scalar field.
Let us start with the true vacuum where the only conical singularity is at the black hole horizon. The conical deficit reads
\begin{align}
\alpha_h = \beta_{\tau_-}/\beta_h \ \ ,
\end{align}
and thus Eq.~\eqref{bulkaction} becomes
\begin{equation}\label{eqn:B_bulk_true}
I_E^{\text{bulk}}(g^-,\varphi_0(T)) = - \frac{A_-}{4 G} + \frac{1}{4 G} \int d\tau_-\, R^2\, \frac{df_-(R)}{dR} + \frac{\beta_{\tau_-}}{4 G} \left( \frac{A_-}{\beta_h} - 2 G M_- + \frac{2 \Lambda_- r_h^3}{3} \right) \ \ .
\end{equation}
In fact, using Eq.~\eqref{Hawtemps}, the last term in brackets vanishes identically. In the false vacuum, we only have the cosmological horizon with a conical deficit
\begin{align}
\alpha_c = \beta_{\tau_+}/\beta_c \ \ ,
\end{align}
and therefore
\begin{equation}\label{eqn:B_bulk_false}
I_E^{\text{bulk}}(g^+, 0) = - \frac{A_c}{4 G} - \frac{1}{4 G} \int d\tau_+\, R^2 \frac{df_+(R)}{dR} + \frac{\beta_{\tau_+}}{4 G} \left( \frac{A_c}{\beta_c} + 2 G M_+ -\frac{2 \Lambda_- r_c^3}{3} \right) \ \ ,
\end{equation}
where similarly the combination in brackets vanishes identically.

Moving on, the surface Euclidean action arises from the Hawking-Gibbons-York boundary term \cite{York:1972sj,Gibbons:1976ue} evaluated at the wall boundary
\begin{equation}
I_E^{\rm surf}(g)=   \frac{1}{8\pi G}\int d^3x\, \sqrt{h}\, \left( K_+ - K_- \right) \ \ ,
\end{equation}
plus the surface energy in the scalar field profile
\begin{equation}
 I_E^{\rm surf}(g,\varphi)=   \sigma \int d^3x\, \sqrt{h} = 4 \pi \sigma \int d\lambda\, R^2(\lambda) \ \ .
\end{equation}
We can combine the two pieces above if we notice that $K_+-K_- =  - 12 \pi G \sigma$ by virtue of Eq.~\eqref{secondjunc}. Hence, the total wall contribution is
\begin{equation}
    I_E^{\rm surf} = - 2\pi \sigma \int d\lambda\, R^2(\lambda) = \frac{1}{2G} \int d\lambda\, R(\lambda) \left( f_+\Dot{\tau}_+ - f_-\Dot{\tau}_- \right) \ \ ,
\end{equation}
where Eq.~\eqref{eqn:ex_curv_diff_comp} has been used to substitute for the surface tension\footnote{This step is justified as long as $\sigma$ is a constant that does not depend on geometry, which is true in our case.}.
The complete bubble action is now easily found by subtracting off the Euclidean action of the false vacuum
\begin{equation}
    I_\textrm{SdS} = -\frac{A_+}{4G} - \frac{A_c}{4G} \ \ ,
\end{equation}
giving
\begin{equation}\label{eqn:B}
B = \frac{A_+ - A_-}{4 G} + \frac{1}{4 G} \int d\lambda\, \left[ \left( 2Rf_+-R^2f_+' \right)\Dot{\tau}_+ - \left( 2Rf_--R^2f_-' \right)\Dot{\tau}_- \right] \ \ .
\end{equation}
This is the general form for the tunneling exponent computed in the thin-wall approximation, and has been obtained in Ref.~\cite{Burda:2015yfa}. The full numerical procedure is now clear; the solution of the equation of motion, Eq.~\eqref{eqn:BH_eom}, which satisfies the temperature matching condition, Eq.~\eqref{eqn:temp_cond}, is obtained, substituted into Eq.~\eqref{eqn:B} and integrated over a single period $\beta_\lambda$. 
These steps are quite involved, nevertheless, the evaluation of $B$ can be simplified somewhat by casting the integral in terms of $R$ rather than $\lambda$. Using $d\lambda=dR\,d\lambda/dR=dR/\sqrt{U(R)}$ and Eq.~\eqref{eqn:juncone} in addition to
\begin{equation}
    (2Rf_\pm-R^2f'_\pm) = 2(R-3GM_\pm) \ \ ,
\end{equation}
the tunneling exponent becomes
\begin{equation}\label{eqn:B_func_R}
B = \frac{A_+-A_-}{4G} + \frac{1}{2G}\int_{R_\text{min}}^{R_\text{max}} dR\ \left[ 2(R-3GM_+) \frac{1}{f_+ \sqrt{U(R)}} \left( f_+-U(R) \right)^{1/2} - \left\{ +\rightarrow - \right\} \right] \ \ .
\end{equation}
This form uncovers the remarkable feature that the tunneling exponent does not {\em explicitly} depend on the exact functional form of the wall trajectory $R(\lambda)$. Rather, it just depends on the potential $U(R)$ and the metric function. This is reminiscent of the typical situation when using the thin-wall approximation, in which the tunneling exponent becomes independent of the exact bubble profile.
Let us also note that it appears as if Eq.~\eqref{eqn:B_func_R} is independent of the temperature of the system. In reality, however, the temperature dependence is manifested in the value of $M^\beta_-$, which is determined by the matching condition of Eq.~\eqref{eqn:temp_cond}. 

In the case of a static solution, Eq.~\eqref{eqn:B} simplifies further. With $R(\lambda)\rightarrow R_s$ and $M_-\rightarrow M_-^s$, Eq.~\eqref{eqn:juncone} becomes
\begin{equation}
    \Dot{\tau}_\pm = \frac{1}{\sqrt{f_\pm(R_s)}} \ \ ,
\end{equation}
and the static bounce action takes the form
\begin{equation}\label{eqn:B_s}
    B_{\rm s} = \frac{A_+-A_-}{4G} + \frac{\beta_\lambda}{2G} \left[ \frac{(R_s - 3GM_+)}{\sqrt{f_+(R_s)}} - \frac{(R_s - 3GM_-^s)}{\sqrt{f_-(R_s)}} \right] \ \ ,
\end{equation}
where it is paramount to realize that $\beta_\lambda$ is not dictated by the dynamics any more since the wall is static. Notice that in the static case it is impossible to change variables as we have done to reach Eq.~\eqref{eqn:B_func_R}, simply because $U(R_{\text{s}}) = 0$, and thus the appropriate result in this case is given by Eq.~\eqref{eqn:B_s}. Since the static solution is the equivalent of the $O(3)$ symmetric configurations in flat-space QFT, it is natural to set $\beta_\lambda = 1/T$. With our mass range, Eq.~\eqref{eqn:M1_range}, the term in square brackets in Eq.~\eqref{eqn:B_s} is negative definite. Finally, it is important to realize that the $\beta_\lambda$-dependent term in Eq.~\eqref{eqn:B_s} is absent from the corresponding action of Refs.~\cite{Gregory:2013hja,Burda:2015isa,Burda:2015yfa,Burda:2016mou} simply because these works only consider vacuum phase transitions. In other words, there is no natural period dictated by the physics and the integral part of Eq.~\eqref{eqn:B} can naturally be set to zero. 

\subsection{Quantitative analysis}
\begin{figure}[t]
    \centering
    \begin{subfigure}{0.59\textwidth}
    \includegraphics[width=\textwidth]{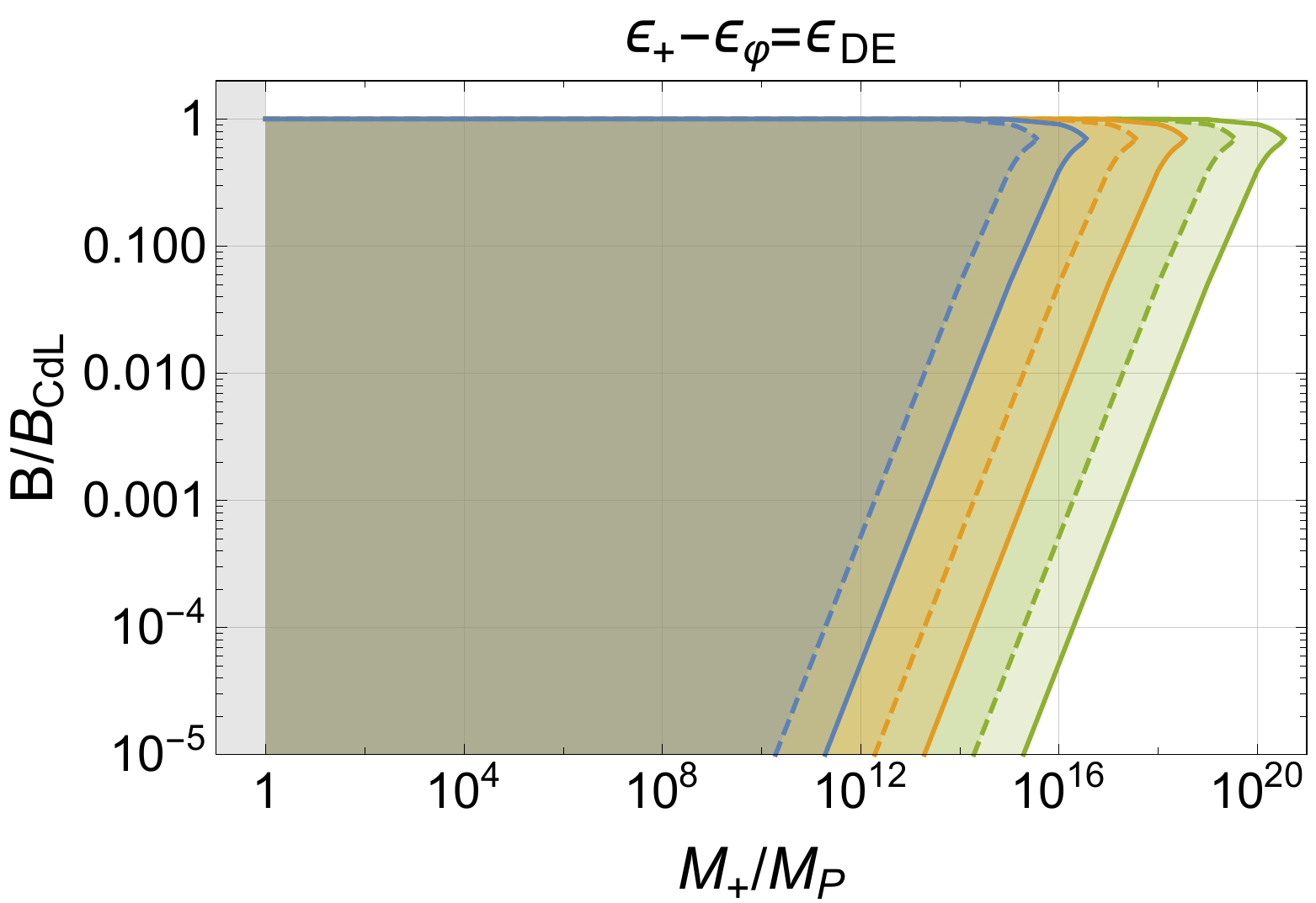}
    \end{subfigure}
    \hfill
    \begin{subfigure}{0.40\textwidth}
    \includegraphics[width=0.9\textwidth]{Figures/plot_legend.pdf}
    \end{subfigure}
    \caption{The tunneling action of Eq.~\eqref{eqn:B} as a function of the seed black hole mass $M_+$, with the upper limit given by $M_+^{\rm max}$. The shaded regions are generated by varying $M_-$ in the admissible range. At fixed $M_+$, the upper limit of any single shaded region represents the point $M_-=M_+$, while the lower limit is the static case $M_-=M_-^s$. Small seed black holes present arbitrary reduction in $B$. For the static solutions an example $T=100\, \textrm{GeV}$ is used. Changing the temperature does not change the generic features.}
    \label{fig:BBCdL_vs_M2_example}
\end{figure}

We are now in a position to quantify the potential effect of black holes on the decay exponent of cosmological first-order phase transitions. To begin, we study the situation in the full range of $M_-$ given by Eq.~\eqref{eqn:M1_range}, without worrying about any matching conditions. As a means to isolate the potential effects of primordial black holes, we compare our tunneling exponent to the CdL result. This is easily obtained by plugging Eq.~\eqref{eq:wallcdl} in Eq.~\eqref{eqn:B} (or Eq.~\eqref{eqn:B_func_R} with $R_\textrm{min}=0$), setting $M_+ = M_- = 0$, and integrating. The result is
\begin{align}\label{eqn:Bcdl}
B_{\rm CdL} &= 2\pi^2\sigma \zeta^{-3/2} + 12\pi^2 M_\textrm{P}^4 \left[ \frac{\left(1-f^{3/2}_+\left(1/\zeta^{1/2}\right)\right)}{\epsilon_+} - \frac{\left(1-f_-^{3/2}\left(1/\zeta^{1/2}\right)\right)}{\epsilon_-}\right] \ \ ,
\end{align}
where $\zeta$ is given in Eq.~\eqref{eq:cdleom}. The results are shown in Fig.~\ref{fig:BBCdL_vs_M2_example}.
With the upper limit given by the maximum remnant mass $M_-=M_+$ and the lower limit by the static solutions $M_-=M_-^s$, the shaded region represents the range of values of $B/B_\textrm{CdL}$ for all solutions. Immediately noticeable is the smooth transition to the CdL limit at the top of any of the shaded regions as the mass approaches zero, i.e. $M_-=M_+\to0$. Compared to $B_\textrm{CdL}$, we observe the possible reduction in $B$ by orders of magnitude, to even an arbitrary degree at low $M_+$. The effect of $M_-$ on $B$ is quite remarkable given how small the mass difference is, as seen in Fig.~\ref{fig:M1_vs_M2}. In fact, the largest reduction in mass results in the greatest reduction in the tunneling exponent, namely, at fixed $M_+$ the smallest $B$ is attained at $M_-^{\rm s}$. Notice that the static branch of solutions is completely disconnected from the CdL limit. 

Additionally, the seed mass $M_+$ plays a crucial role. Increasing $M_+$ drastically pushes up the lower limit of $B$, thus limiting the possible reduction. On the other hand, effects on the upper limit of $B$ are less pronounced as we increase $M_+$. We do not observe any noticeable difference up until the maximum point, $M_+^\textrm{max}$, at which $B$ starts to slightly decrease. As expected, at $M_+^\textrm{max}$ both the upper and lower limits on $B$ meet at the same point. The surprising feature is that regardless of all the other parameters, the ratio $B/B_\textrm{CdL}$ attains a common value of $\sim 0.7$. Finally, the effects of $\sigma$ and $\epsilon_\varphi$ are simple yet substantial; by changing the maximum $M_+^\textrm{max}$, as in Fig.~\ref{fig:M2Crit_vs_e2}, the profile of $B$ is shifted correspondingly. In particular, at fixed value of $B/B_\textrm{CdL}$ the largest accessible seed mass changes by orders of magnitude by varying $\sigma$ and $\epsilon_\varphi$. To summarise, transitions with the largest change in black hole mass, culminating in static solutions, possess the greatest prospects for improving transition rates. This improvement is reduced as the seed black hole mass $M_+$ is increased.

We end this section by comparing our findings to the main conclusions drawn up in Refs.~\cite{Gregory:2013hja,Burda:2015isa,Burda:2015yfa,Burda:2016mou}, although it is important to stress that we consider finite-temperature phase transitions in contrast to the vacuum case studied in those references. Yet, it is still true that our conclusions about the dominant tunneling configurations are in exact agreement with the findings of those papers. In particular, Ref.~\cite{Burda:2015yfa} found that the dominant tunneling configuration is either a static solution or an oscillating solution but without a remnant black hole ($M_-=0$), see Fig.~9 in appendix A of Ref.~\cite{Burda:2015yfa}. The delineation between the two situations, at fixed seed mass $M_+$, is dictated by the factor $\bar{\sigma} \ell$ in Refs.~\cite{Gregory:2013hja,Burda:2015isa,Burda:2015yfa,Burda:2016mou}. In our case this factor is $\bar{\sigma} \ell \simeq \sigma \sqrt{\epsilon_\varphi}/ M_P$, which for all our parameter space is a very tiny number. As evident from the analysis in Ref.~\cite{Burda:2015yfa}, as the factor $\bar{\sigma} \ell$ approaches zero the dominant solution is just given by the static branch. This is precisely in agreement with what we observe in Fig.~\ref{fig:BBCdL_vs_M2_example} above.

\subsection{Bubble nucleation criteria with black holes}\label{sec:nuc_criteria}
Before we discuss an example electroweak phase transition, it is important to define the nucleation criteria in the presence of black holes. In a typical (flat-space) first-order cosmological phase transition proceeding through thermal excitation, the transition rate per unit volume is given by the expression
\begin{equation}\label{eqn:flat_rate}
    \frac{\Gamma}{V} = A\, e^{-B_\textrm{flat}}\quad ,\quad B_\textrm{flat}=\frac{S_3}{T}\ ,
\end{equation}
where $S_3$ is the energy of the critical bubble and the prefactor $A$ is of mass dimension four, commonly approximated as $T^4$. A successful transition is defined to be the nucleation of one bubble per Hubble time per Hubble volume. In a radiation dominated Universe, this results in a nucleation condition on the exponent
\begin{equation}
    B_\textrm{flat}^\textrm{nuc.}\sim 4\log\left( \frac{M_P}{T} \right)\ .
\end{equation}
If the exponent ever reaches this value or below the transition is deemed successful. For electroweak-scale transitions this takes the value $B_\textrm{flat}\sim140$.

Once black holes are included, however, one cannot define a nucleation rate per volume because the presence of the black hole breaks the spatial translation symmetry of the instanton. Therefore, there is no factor of volume, as in Eq.~\eqref{eqn:flat_rate}, and instead we have a transition rate
\begin{equation}\label{eqn:BH_rate}
    \Gamma = A\, e^{-B}\ ,
\end{equation}
describing nucleation around a single black hole, where the prefactor $A$ is of mass dimension one. Unfortunately, the prefactor $A$ is unknown. Even in the simple case where gravity is assumed {\em not} to be quantized, computing the determinant of one-loop fluctuations of the scalar field requires massive work since the background geometry is very involved. Still, following Ref.~\cite{Gregory:2013hja}, one can attempt a rough estimate using $A\sim (GM_+)^{-1}$. An alternative nucleation condition can then be derived for the black hole case. Requiring that a single bubble be nucleated per Hubble time, the nucleation condition on $B$ is now
\begin{equation}\label{eqn:BH_nuc}
    B_\textrm{BH}^\textrm{nuc.} \sim \log\left( \frac{M_P^3}{T^2 M_+} \right)\ .
\end{equation}
For an electroweak-scale transition and a reference mass $M_+=10^{14}\, M_P$ this condition is $B_\textrm{BH}\sim 42$. One then might be tempted to think that black holes are not efficient in the nucleation process because a lower $B$ is harder to achieve in general. However, given the large reductions possible in Euclidean action seen in Fig.~\ref{fig:BBCdL_vs_M2_example}, it is likely black holes can satisfy this condition, Eq.~\eqref{eqn:BH_nuc}, at a temperature higher than the usual temperatures required for flat space, thereby dominating the nucleation process.

As mentioned, Eq.~\eqref{eqn:BH_rate} describes the rate of nucleation around a single black hole. In reality, there will be a population of primordial black holes with a given number and mass distribution. To get an idea of the true nucleation rate, Eq.~\eqref{eqn:BH_rate} should be multiplied by the total number of black holes $N_\textrm{PBH}$ in a given Hubble volume. Hence, the nucleation condition, now the probability to nucleate one bubble per Hubble time across the full population of black holes, reads
\begin{equation}\label{eqn:BH_nuc_pop}
    B_\textrm{BH}^\textrm{nuc.} \sim \log\left( \frac{N_\textrm{PBH} M_P^3}{T^2 M_+} \right)\ .
\end{equation}
A larger number of black holes, $N_\textrm{PBH}$, will increase $B_\textrm{BH}^\textrm{nuc.}$ therefore making nucleation easier to achieve, as one would expect.

\subsection{Electroweak-like phase transition}\label{sect:ewfinal}
As we explained in Sect.~\ref{sec:BH_and_PT}, there are only two tunneling configurations at any fixed value of the parameters $(M_+,\epsilon_+,\epsilon_\varphi,\sigma)$. These are parametrized by $M_-^\beta$ and $M_-^{\rm s}$, which represent the oscillating and static instantons of Eq.~\eqref{eqn:BH_eom}. It is crucial to note that the matched solution, $M_-^\beta$, is not guaranteed to exist because the period, $\beta_\lambda$, might never be equal to the inverse temperature of the system for $T < T_c$. If the two solutions exist, the decay rate will then be determined by the smaller tunneling exponent amongst both configurations. Our goal now is to compare the tunneling exponents of the two solutions. Indeed, Fig.~\ref{fig:BBCdL_vs_M2_example} shows that the static solution, at fixed $M_+$, has the lowest action, nevertheless, we aim to know the exact difference in the tunneling exponent when static solutions are compared with their oscillating counterparts which satisfy the condition Eq.~\eqref{eqn:temp_cond}.

We have in mind an electroweak-like scenario, where tunneling proceeds via the following finite-temperature potential 
\begin{equation}\label{eqn:EW_pot}
    V(\varphi, T) = \frac{1}{2}\left( DT^2-\mu^2 \right)\varphi^2 - \frac{E}{3}T\varphi^3 + \frac{\lambda}{4}\varphi^4 \ \ ,
\end{equation}
where, for definiteness, $\mu = 88\, \textrm{GeV}$ and $\lambda=0.129$. The values of $D$ and $E$ are dependent on the details of the underlying model. We use the standard model value of $D=0.34$. To replicate BSM effects and get a stronger first-order phase transition\footnote{Such an effect can easily be achieved by, for example, adding a scalar singlet to the standard model that does not gain a vacuum-expectation-value.}, the value of $E$ is enhanced above its standard model value and taken to be $E=0.21$. Consequently, we have a critical temperature $T_c=171.3\, \textrm{GeV}$. The surface tension is calculated using the equation 
\begin{equation}\label{eq:sigma}
    \sigma = \int_{\varphi_+}^{\varphi_-} d\varphi\, \sqrt{2V(\varphi,T_c)}\ ,
\end{equation}
giving the value $\sigma=2.2\times10^5\, \textrm{GeV}^3$. Finally, $\epsilon_\varphi$ is a function of temperature that we do not quote here. A comment is due at this stage, the formula for the surface tension in Eq.~\eqref{eq:sigma} is only valid in flat space. In the present case, one has to double check if the radius of the bubble is large enough compared to the horizon size of the remnant black hole. In our parameter space, we found that this is indeed true, and so we continue to use the simple formula in Eq.~\eqref{eq:sigma}.

\begin{figure}[t]
    \centering
    \includegraphics[width=0.59\textwidth]{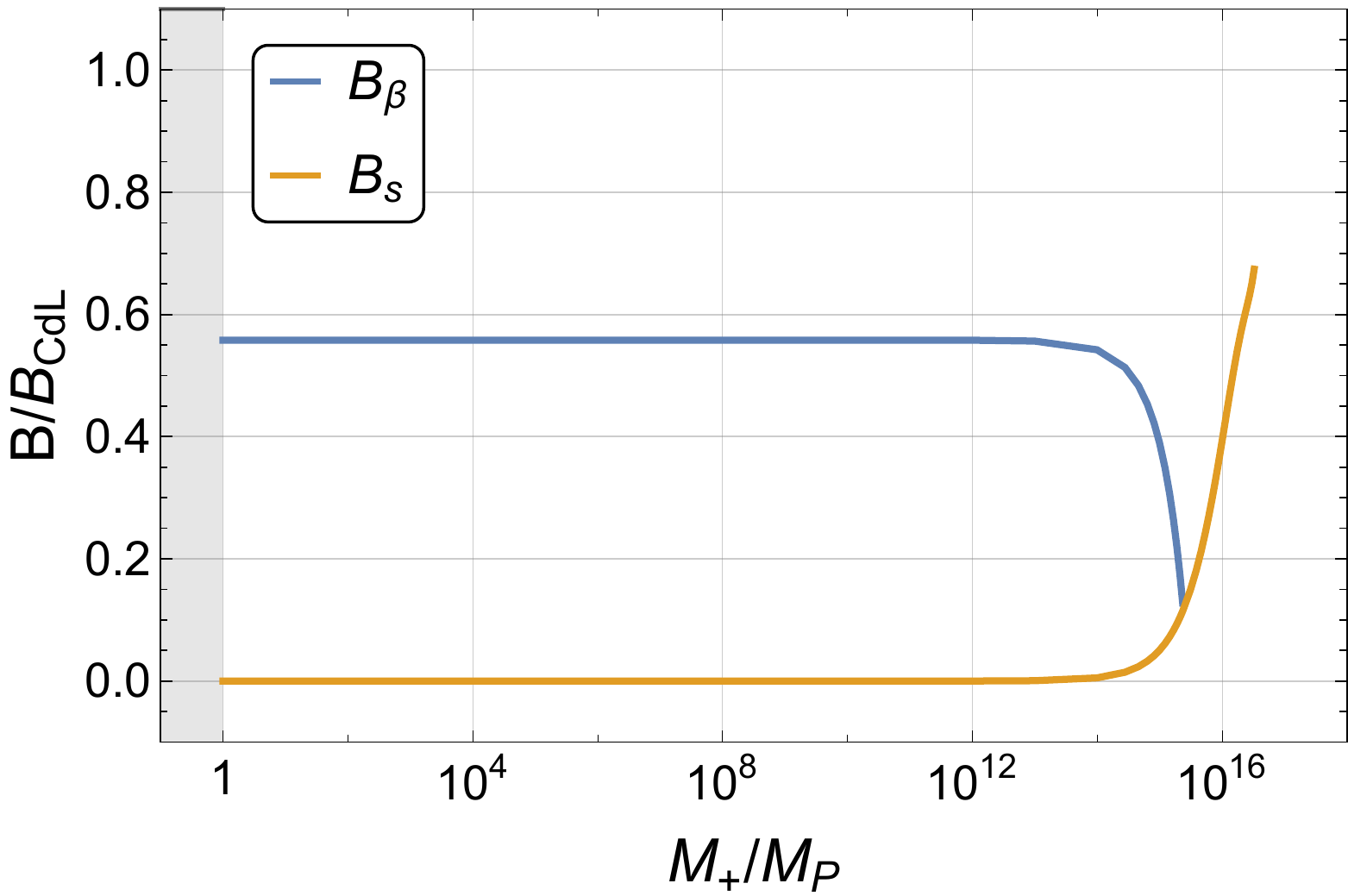}
    \caption{The tunneling action as a function of the seed black hole mass $M_+$ for some example electroweak-scale parameters $T_c=112\, \textrm{GeV}$ and $\sigma=10^4\, \textrm{GeV}^3$. Blue Line: Matching oscillating solutions calculated using Eq.~\eqref{eqn:B_func_R}. Orange Line: Static solutions calculated using Eq.~\eqref{eqn:B_s}. Matching occurs at $T=111.4\, \textrm{GeV}$ where $\epsilon_\varphi=10^7\, \textrm{GeV}^4$.}
    \label{fig:B_match}
\end{figure}

Unfortunately, with these typical parameters we did not find it possible to obtain a valid oscillating solution which hinders the comparison that we aim for. We circumvent this by using $T_c=112.8\, \textrm{GeV}$, $\sigma=10^4\, \textrm{GeV}^3$ and a mock $\epsilon_\varphi$. We have scanned the range $10^6\leq\epsilon_\varphi\leq 10^8\, \textrm{GeV}^4$ to find an oscillating solution satisfying Eq.~\eqref{eqn:temp_cond}. Fig.~\ref{fig:T_vs_M2} shows that, in comparison to $M_+$, small changes in $\epsilon_\varphi$ strongly alter the inverse period of the oscillating solutions; there is likely only a small window over which matching is possible. For our example matching is satisfied at $T=111.4\, \textrm{GeV}$ and $\epsilon_\varphi=10^7\, \textrm{GeV}^4$. The tunneling exponent of both solutions are found and compared in Fig.~\ref{fig:B_match}.
We observe that the oscillating solution provides a significant reduction in $B$, however, static solutions always dominate as expected. Increasing $M_+$ has the opposite effect on the two solutions. While both remain essentially constant over small values of the seed mass, at higher $M_+$ the oscillating result $B_\beta$ drops rapidly while the static result $B_{\rm s}$ rises. The increase in $B_{\rm s}$ can already be seen from Fig.~\ref{fig:BBCdL_vs_M2_example}. The drop in $B_\beta$ is due to the matching remnant mass, $M_-^\beta$, approaching the static limit $M_-^{\rm s}$ as $M_+$ increases, as illustrated by Fig.~\ref{fig:M1_vs_M2}. The maximum value of $M_+$ where matching is still attainable is definitely smaller than the absolute maximum $M_+^\textrm{max}$ and represents the point where the two curves meet in Fig.~\ref{fig:B_match}. To be clear, we observe that the lines never cross and the static solutions are always dominant over the periodic solutions.

Finally, we turn back to the realistic scenario of Eq.~\eqref{eqn:EW_pot} and focus entirely on the static solutions which, as explained, provide the dominant tunneling configurations. In fact, our central equation~\eqref{eqn:B_s} can be simplified further to highlight its thermodynamic meaning. With the values $\sigma\simeq 2.7\times10^5\, \textrm{GeV}^3$ and $\epsilon_\varphi\simeq10^{6}\, \textrm{GeV}^4$, notice that we possess a hierarchy $r_h\ll R_s \ll r_c$, where $r_h$ and $r_c$ are the black hole and cosmological horizon radii respectively (see Fig.~\ref{fig:Rs_vs_M2}). In this case, $f_\pm(R_s)\simeq1$ and therefore the static tunneling action becomes
\begin{equation}
    B_s \simeq \frac{A_+-A_-}{4G} - \frac{3\beta}{2} (M_+-M^s_-)\ .
\end{equation}
The meaning of each term is now transparent. The first is the difference in Hawking entropy between the seed and remnant black holes. While the second represents the difference between the Arnowitt-Deser-Misner (ADM) masses (or the energy) of seed and remnant black holes.

\begin{figure}[t]
     \centering
     \begin{subfigure}{\textwidth}
     \centering
     \includegraphics[width=0.59\textwidth]{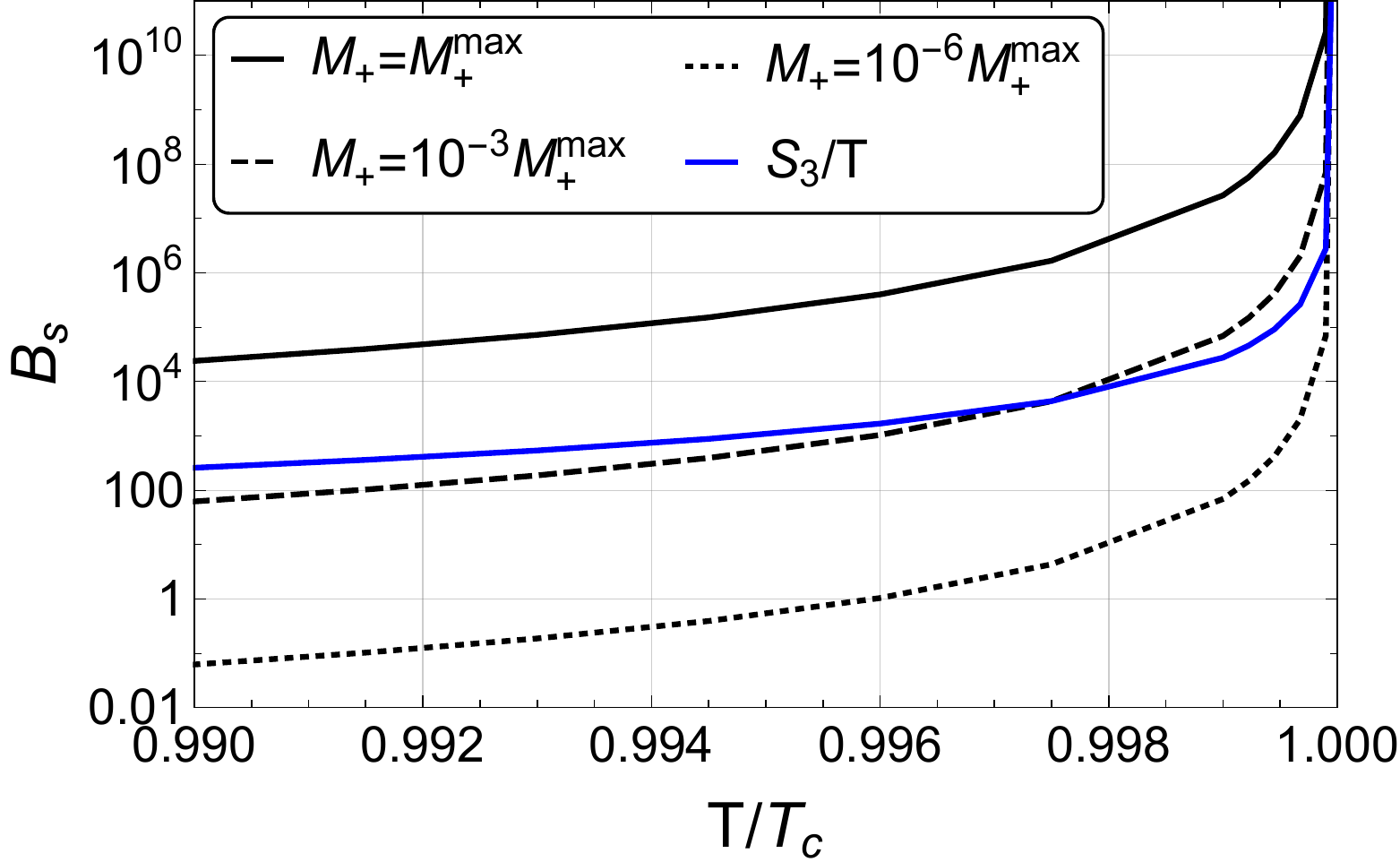}
     \end{subfigure}
     \vfill
     \vspace{1em}
     \begin{subfigure}{\textwidth}
     \centering
     \includegraphics[width=0.59\textwidth]{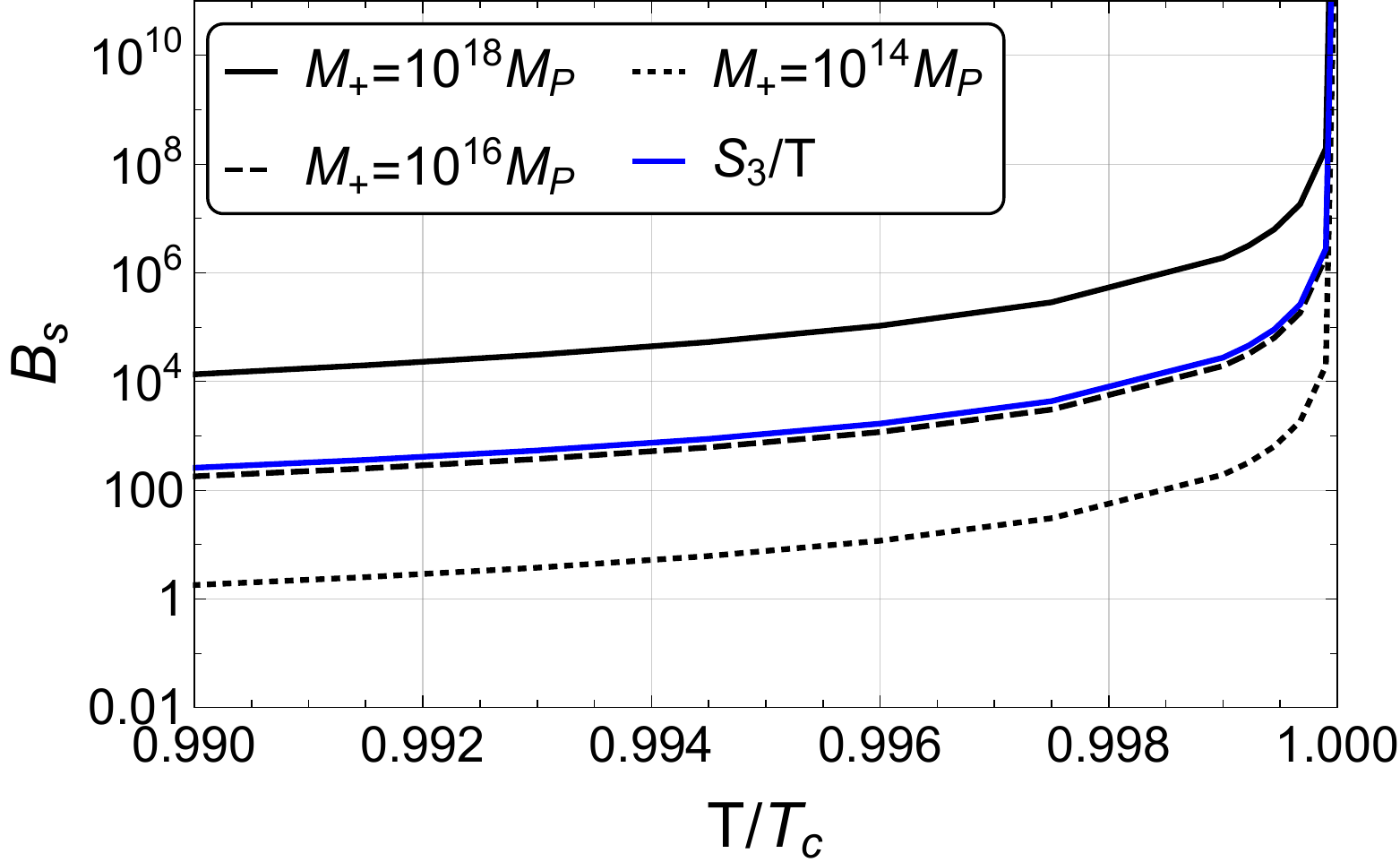}
     \end{subfigure}
     \caption{The static Euclidean action for black hole induced tunneling $B_\textrm{s}$ (black lines) and the typical thermal tunneling rate $S_3/T$ (blue line) for the example electroweak potential of Eq.~\eqref{eqn:EW_pot}, with the critical temperature $T_c=171.3\, \textrm{GeV}$ and surface tension $\sigma=2.7\times 10^5\, \textrm{GeV}^3$. Note that $M_+^\textrm{max}$ decreases with $T$ due to increasing $\epsilon_\varphi$. Top: $M_+$ is a fraction of $M_+^\textrm{max}$. The solid, dashed and dotted lines represent $M_+=M_+^\textrm{max},\, 10^{-3}M_+^\textrm{max}$ and $10^{-6}M_+^\textrm{max}$ respectively. Bottom: $M_+$ is a constant value. The solid, dashed and dotted lines represent $M_+=10^{18}M_\textrm{P},\, 10^{16}M_\textrm{P}$ and $10^{14}M_\textrm{P}$ respectively.}
     \label{fig:B_vs_T}
\end{figure}

Now we can ask, how efficient are black holes in seeding the transition? To get an idea, we compare $B_{\rm s}$ to the thermal tunneling exponent in flat-space given by the familiar expression
\begin{align}
B_{\rm flat} = \frac{S_3}{T} \ \ ,
\end{align}
where $S_3$ denotes the energy of the $O(3)$ invariant critical bubble\footnote{We note that the comparison between $B_{\rm s}$ and $B_{\rm CdL}$ is not really meaningful for phenomenology, and this is why we employ $S_3/T$ instead. This is because the CdL result represents decay via quantum vacuum fluctuations ($T$=0), and is not appropriate for thermal transitions. Ideally, one might try and compute the thermal tunneling exponent in the mere presence of a cosmological constant, and compare directly to $B_{\rm s}$. Nevertheless, we believe this will not be important since, for phenomenological purposes, $B_{\rm CdL}$ is essentially identical to the flat-space $O(4)$-invariant bounce action~\cite{Coleman:1980aw}.}. Fig.~\ref{fig:B_vs_T} shows the result which exhibits interesting features that might not be expected a priori.
First of all, the presence of black holes does not necessarily improve the tunneling rate. For example, if we use the maximum available value, $M_+^{\rm max}$, no reduction is observed. Second, one has to go to relatively lower values of the seed mass in order to attain a noticeable reduction. The reduction in $B$, in and of itself, is a positive outcome but, as outlined in Sect.~\ref{sec:nuc_criteria}, the two scenarios have different nucleation conditions. Hence, a comparison between $B_s$ and $S_3/T$ is not strictly correct. Rather, the relevant comparison is that of the respective nucleation temperatures. However, even in the most restricted case of a single black hole (where the nucleation condition is $B_s\lesssim 42$ as found from Eq.~\eqref{eqn:BH_nuc}), Fig.~\ref{fig:B_vs_T} shows that black hole initiated nucleation occurs before traditional thermal tunneling for $M_+\lesssim10^{15}\, M_P$. Prospects for black holes improving transition rates are certainly very promising. 

The complete picture, however, can only be understood after some further considerations. Firstly, various aspects, which will be outlined in Sect.~\ref{sec:pheno}, constrain the phenomenological viability of seed black hole masses at formation. Secondly, knowledge of the precise mass and number distributions of PBHs is necessary to make accurate predictions in regards to the way the transition proceeds. While consideration of a specific PBH population is left to future work, we discuss the main transition scenarios in Sect.~\ref{sec:discussion}.

\section{Routes for phenomenology}\label{sec:pheno}
We have provided a framework for calculating the effect of black holes on generic finite temperature first-order phase transitions, encompassing arbitrary masses, cosmological constants and temperatures. Our central results for an EW-like transition, displayed in Fig.~\ref{fig:B_vs_T}, need a few more inputs in order to conduct a full phenomenological study and assess accurately the role of primordial black holes in cosmological phase transitions. This section is a presentation of what inputs we require to achieve our goal in future works. 

\subsection{Primordial Black Holes: Formation and Abundance}\label{sec:PBHs}
The first consideration needed for phenomenology is the issue of primordial black holes, in particular, their mass spectrum and abundance close to the electroweak epoch. PBHs can be formed through various mechanisms~\cite{Carr:1974nx,Crawford:1982yz,Hawking:1982ga,Hawking:1987bn} (see reviews~\cite{Green:2014faa,Carr:2020gox}). For instance, in the case formation proceeds due to the collapse of large density perturbations during the radiation era~\cite{Carr:1974nx,Carr:1975qj,Niemeyer:1997mt,Niemeyer:1999ak,Green:2004wb,Young:2014ana}, an estimate of the mass is given by~\cite{Green:2014faa}
\begin{align}\label{eqn:mpbh}
    M_\textrm{PBH} \simeq \frac{c^3 t}{G}
    &\simeq 10^{43}\, \left(\frac{t_{\rm cos}}{\text{sec.}}\right)\, M_{\rm P} \ \ ,
\end{align}
where $t_{\rm cos}$ denotes cosmological time. Therefore, a wide range of PBH masses becomes available between the end of inflation and the EW epoch. For instance, shortly following inflation, $t_{\rm cos}\sim 10^{-32}\, \textrm{seconds}$ and the approximate mass is $M_\textrm{PBH}\simeq 10^{10}M_P$, while a PBH forming at the EW epoch is as large as $M_\textrm{PBH}\simeq 10^{30}M_P$. Given the central results of Fig.~\ref{fig:B_vs_T}, we are guaranteed to find relevant PBH masses, $M_\textrm{PBH}\lesssim 10^{15} M_P$, at the EW epoch. 

Additionally, the observational constraints on PBH abundance need to be taken into account. For the mass range we are interested in, the most stringent constraint derives from Big Bang Nucleosynthesis (BBN) because such black holes would have evaporated at the epoch of BBN (see Fig.~4 in Ref.~\cite{Carr:2020gox}). Precisely, black holes in the mass range $\sim10^{15}-10^{19} M_P$ can comprise between $\sim10^{-17} - 10^{-24}$ of the fraction of the Universe's energy density at the time of their formation~\cite{Carr:2020gox}. Therefore, for a given relevant seed mass $M_+$, and knowing the relation between formation time and mass, Eq.~\eqref{eqn:mpbh}, we can directly constrain the available number density at the EW epoch\footnote{Ref.~\cite{Dai:2019eei} considered the dependence of the PBH masses and number density, at formation time, on the spectral index of primordial density fluctuations.}. The direct consequence of this bound concerns how the phase transition actually proceeds, as we discuss in detail in Sect.~\ref{sec:discussion}.

\subsection{Black Hole Decay via Hawking Radiation}
It is well known that black holes decay via Hawking radiation~\cite{Hawking:1974rv,Hawking:1974sw}, endangering the completion of phase transitions seeded by black holes\footnote{A recent paper~\cite{Hayashi:2020ocn} performed an interesting study that looks in detail at the effect of Hawking radiation on the dynamics of a \emph{vacuum} transition, considered previously in Refs.~\cite{Gregory:2013hja,Burda:2015isa,Burda:2015yfa,Burda:2016mou}.}. To claim a black hole of certain mass $M_+$ is relevant to the phase transition, at least one bubble per horizon volume must nucleate \emph{before} the black hole decays via Hawking radiation. Therefore, the transition rate $\Gamma$, given in Eq.~\eqref{eqn:BH_rate}, must be greater than the Hawking evaporation rate $\Gamma_H$ \cite{Gregory:2013hja}. However, calculating $\Gamma$ requires knowledge of the coefficient $A$. As discussed in Sect.~\ref{sec:nuc_criteria}, this is unknown once gravitational effects are included and, at best, a very rough bound on the seed mass can thus be made.

Given the known nucleation timescale of EW-like phase transitions in the absence of black holes, we can adopt a worst case scenario to estimate a lower bound on seed masses. Our logic is as follows; we are only interested in the scenario where black holes noticeably improve the transition rate. This then dictates the \emph{worst case} transition timescale to be that of flat space, $\tau_\textrm{EW}\sim 10^{-11}$ seconds. Given this strategy, we use the known lifetime of a black hole, of mass $M$, against Hawking evaporation (see e.g.~\cite{Carr:2009jm}) to place a lower bound on $M_+$. This translates to the condition
\begin{equation}\label{eq:Htau}
    M_+ \gtrsim \left( \frac{\tau_\textrm{EW}}{G^2} \right)^{1/3} \simeq 10^{12} M_P \ \ .
\end{equation}
We remind that this is not a very strict bound because black holes of such masses, as we have shown, enhance the tunneling rate appreciably which further loosens the bound on $M_+$ from Hawking evaporation. Finally, we stress again that it is important to compute, or at least properly estimate, the coefficient $A$ in order to obtain decisive bounds.

\section{Discussion and prospects for gravitational waves}\label{sec:discussion}
In light of the constraints outlined in Sect.~\ref{sec:pheno}, we return to our electroweak-like example, Eq.~\eqref{eqn:EW_pot}, and discuss the consequences for gravitational wave production. Given that the dependence on temperature is virtually the same as in flat-space, more general conclusions can be made about the nature of the phase transition. It is clear that the interplay between a black hole seeded transition and traditional thermal excitation offers a variety of physical scenarios for nucleation. This is strictly dependent on both the seed mass $M_+$ as well as the number density of primordial black holes, $n_\textrm{PBH}$, at the EW epoch. This latter quantity is indeed constrained by BBN observations, as we discussed. In Sect.~\ref{sec:nuc_criteria} we derived how the condition on the Euclidean action for successful nucleation differs between the traditional case, $B_\textrm{flat}^\textrm{nuc.}$, and the black hole case, $B_\textrm{BH}^\textrm{nuc.}$. Using these conditions, and assuming that the corresponding Euclidean actions $B_s$ and $S_3/T$ indeed have the same dependence on $T$ (as observed in Fig.~\ref{fig:B_vs_T}), we now describe these possible scenarios and the consequences for the resulting gravitational wave spectrum~\cite{Caprini:2015zlo,Caprini:2019egz}. 
\paragraph{Typical thermal excitation dominates:} First is the simple case where black holes play no role and the traditional thermal excitation dominates. In terms of the nucleation conditions, that is $B_s\gg B_\textrm{BH}^\textrm{nuc.}$ and $S_3/T \leq B_\textrm{flat}^\textrm{nuc.}$. Such a scenario is most likely at large seed masses $M_+$.
\paragraph{Black holes dominate, many black holes:} The second scenario is the opposite; there are many black holes and the transition proceeds solely via nucleation around them. This corresponds to a large $n_\textrm{PBH}$ with $B_s\leq B_\textrm{BH}^\textrm{nuc.}$ and $S_3/T\gg B_\textrm{flat}^\textrm{nuc.}$. Consequently, the bubble properties are heavily influenced by the distribution of black holes - a large number of black holes, and therefore nucleation sites, results in smaller bubbles at collision. Hence, a gravitational wave spectrum with a higher frequency and reduced amplitude is produced.
\paragraph{Black holes dominate, few black holes:} The third case is again a transition seeded exclusively by black holes but now with a small number of black holes. That is small $n_\textrm{PBH}$ with $B_s< B_\textrm{BH}^\textrm{nuc.}$ and $S_3/T\gg B_\textrm{flat}^\textrm{nuc.}$. Fewer bubbles are nucleated and they therefore have a larger radius upon collision. The result could be a strong gravitational wave signal at lower frequencies.
\paragraph{Mixed:} Lastly, there could be a crossover scenario where both typical thermal excitation and black hole nucleation can occur. That is $B_s\sim B_\textrm{BH}^\textrm{nuc.}$ and $S_3/T\sim B_\textrm{flat}^\textrm{nuc.}$. The outcome here is again highly dependent on the distribution of black holes. If there are few, the typical thermal transition is likely to dominate and proceed as normal. If there are many, then black hole nucleation could dominate resulting in many, smaller bubbles. As previously mentioned, this produces a gravitational wave spectrum with a higher frequency and reduced amplitude.\\

Having described the possible scenarios, we can now reconsider our basic EW-like transition of Sect.~\ref{sect:ewfinal}, described by Eq.~\eqref{eqn:EW_pot}. Clearly, with our values of the coefficients $D$ and $E$ it is true that $S_3/T\sim B_\textrm{flat}^\textrm{nuc.}$ and therefore we are in the mixed case. It remains then to decide if black holes can dominate the nucleation process. This requires that we know both the average number of bubbles with a flat space core and the average number of black holes $N_\textrm{PBH}$ in the mass range $10^{12}-10^{15} M_P$.

The average number of bubbles with a flat space core can be calculated quite easily knowing the average bubble radius. Using Ref.~\cite{Caprini:2019egz}, we get $N_\textrm{bubbles}\simeq10^{10}$ (this number is based on using a bubble wall velocity of $v_\textrm{wall}=1/3$). On the other hand, the BBN constraints can be used directly to set an upper bound on the total number of black holes per Hubble volume at the EW epoch, given the relevant mass range $10^{12}-10^{15} M_P$. The quantity constrained by BBN is the fraction of the Universe’s mass in PBHs at their formation time~\cite{Carr:2020gox},
\begin{equation}
    \beta(M) \equiv \frac{\rho_\textrm{PBH}(t_\textrm{form})}{\rho(t_\textrm{form})} \simeq \frac{M}{T_\textrm{form}} \frac{n_\textrm{PBH}(t)}{s(t)}\ ,
\end{equation}
where $s(t)$ is the entropy density and the ratio $n_\textrm{PBH}/s$ is conserved. For our mass range, we read off the bound $\beta(M)\lesssim10^{-17}$ from Ref.~\cite{Carr:2020gox}. Using Eq.~\eqref{eqn:mpbh} to relate the formation time to the mass, we find simply
\begin{equation}
    N_\textrm{PBH}\lesssim10^{13}\ .
\end{equation}
We observe the total number of black holes could be orders of magnitude larger than the average number of bubbles with a flat space core, $N_\textrm{bubbles}$. This means there is a real possibility for black hole seeds to dominate the transition. Notwithstanding, one still needs a dedicated study to simulate the seeded bubble collisions in order to predict the resulting gravitational wave spectrum.

\section{Summary}\label{sec:summary}

In this study we have provided a basis for quantifying the effect of black holes acting as nucleation sites for bubbles during a finite-temperature cosmological phase transition. Using a thin-wall formalism, the equation of motion describing the bubble wall is derived building on earlier works~\cite{Hiscock:1987hn,Gregory:2013hja}. There are two types of solution - oscillating and static. By comparing to the typical flat-space solutions, it is clear that these correspond to thermally-assisted quantum tunneling and thermal excitation respectively. While a static solution is always available, an oscillating solution is only valid when the inverse of its period, $\beta_\lambda^{-1}$, matches the temperature, $T$, of the Universe. 

We then calculate the transition exponent, given by the Euclidean action, including the conical singularities arising from a mismatch between the Hawking temperature of the black holes and the periods of Euclidean time. It turns out that the static solutions are always dominant (have the lowest action). The static action is given in Eq.~\eqref{eqn:B_s} and simply requires four input parameters: the seed black hole mass $M_+$, false vacuum energy density $\epsilon_+$, bubble surface tension $\sigma$ and the change in vacuum energy density $\epsilon_\varphi$. While the last two parameters, $\sigma$ and $\epsilon_\varphi$, are provided by the scalar field theory describing the transition, $M_+$ and $\epsilon_+$ are free parameters. A quantitative analysis across the parameter space revealed that reducing the seed mass $M_+$ reduces the action, becoming arbitrarily small at lower masses.

While a reduction in transition exponent is promising for improving transition rates, it does not tell the full story - a comparison to typical thermal excitation in flat-space must be made. In particular, the nucleation temperatures of the two methods must be compared. Importantly, the criterion for successful nucleation is altered in the presence of black holes and this new form is given in Eq.~\eqref{eqn:BH_nuc_pop}. To provide a realistic example, the formalism was applied to an electroweak-like phase transition described by Eq.~\eqref{eqn:EW_pot}. Focusing on the dominant static solutions, Fig.~\ref{fig:B_vs_T} shows that although black hole seeds do not always improve nucleation rates (larger seed masses), enhancements are observed for seed masses $M_+\lesssim 10^{15}\, M_P$.

It is clear that black holes acting a nucleation sites could have significant consequences for finite temperature cosmological phase transitions, offering improvements to transition rates by greatly reducing the transition exponent. In Sect.~\ref{sec:discussion}, we outlined the various transition scenarios based on a comparison between the two tunneling mechanisms which are expected to coexist. For our EW example, we found that BBN constraints are not too stringent and indeed black holes can play a dominant role in cosmological phase transitions.

\section*{Acknowledgements}
We would like to thank Chris Byrnes for an insightful discussion. BKE is supported by UK STFC Consolidated Grant ST/P000819/1. SJH is supported in part by UK STFC Consolidated Grant ST/P000819/1. JPM is supported by a PhD studentship jointly funded by STFC and the School of Mathematical
and Physical Sciences of the University of Sussex.

\bibliographystyle{utphys}
\bibliography{BH_Seeds}

\end{document}